\documentclass[
 amsmath,amssymb,
 aps,
]{revtex4-2}

\usepackage{graphicx}% Include figure files
\usepackage{dcolumn}% Align table columns on decimal point
\usepackage{bm}% bold math
\usepackage{caption}
\usepackage{subcaption}
\usepackage{amsmath}
\usepackage{color}
\usepackage{siunitx}
\usepackage{hyperref}% add hypertext capabilities
\begin{document}

\preprint{APS/123-QED}

\title{Interaction between low-level jets and wind farms\\ in a stable atmospheric boundary layer}% Force line breaks with \\

\author{Srinidhi N. Gadde}
 \email{s.nagaradagadde@utwente.nl}
 \altaffiliation{Physics of Fluids Group, Max Planck Center Twente for Complex Fluid Dynamics, J. M. Burgers Center for Fluid Dynamics and MESA+ Research Institute, University of Twente, P. O. Box 217, 7500 AE Enschede, The Netherlands}%Lines break automatically or can be forced with \\
\author{R. J. A. M. Stevens}%
\affiliation{%
Physics of Fluids Group, Max Planck Center Twente for Complex Fluid Dynamics, J. M. Burgers Center for Fluid Dynamics and MESA+ Research Institute, University of Twente, P. O. Box 217, 7500 AE Enschede, The Netherlands
}%%

\date{\today}% It is always \today, today,
 % but any date may be explicitly specified

\begin{abstract}
{\color{black} 
Low-level jets (LLJs) are the wind maxima in the lower regions of the atmosphere with a high wind energy potential. Here we use large-eddy simulations to study the effect of LLJ height on the flow dynamics in a wind farm with $10\times4$ turbines. We change the LLJ height and atmospheric thermal stratification by varying the surface cooling rate. We find that the first row power production is higher in the presence of a LLJ compared to a neutral reference case without LLJ. Besides, we show that the first row power production increases with decreasing LLJ height. Due to the higher turbulence intensity, the wind turbine wakes recover faster in a neutral boundary layer than in a stably stratified one. However, for strong thermal stratification with a low-height LLJ, the wake recovery can be faster than for the neutral reference case as energy can be entrained from the LLJ. Flow visualizations reveal that under stable stratification the growth of wind farm's internal boundary layer is restricted and the wind flows around the wind farm. Wind farms extract energy from LLJs through wake meandering and turbulent entrainment depending on the LLJ height. Both effects are advantageous for wake recovery, which is beneficial for the performance of downwind turbines. This finding is confirmed by an energy budget analysis, which reveals a significant increase in the kinetic energy flux in the presence of a LLJ. The jet strength reduces as it passes through consecutive turbine rows. For strong stratification, the combined effect of buoyancy destruction and turbulence dissipation is larger than the turbulent entrainment. Therefore, the power production of turbines in the back of the wind farm is relatively low for strong atmospheric stratifications. We find that the pronounced wind veer in stably stratified boundary layers creates asymmetry in the available wind resource, which can only be studied in finite-size wind farm simulations. We emphasize that spanwise-infinite wind farm simulations may underpredict wind farm performance as the additional beneficial effect of LLJ cannot be observed.}
\end{abstract}

\maketitle

%\tableofcontents

\section{\label{sec1}Introduction}
\begin{figure}
 \centering
 \begin{subfigure}[h]{1.0\textwidth}
 \includegraphics[width=\linewidth]{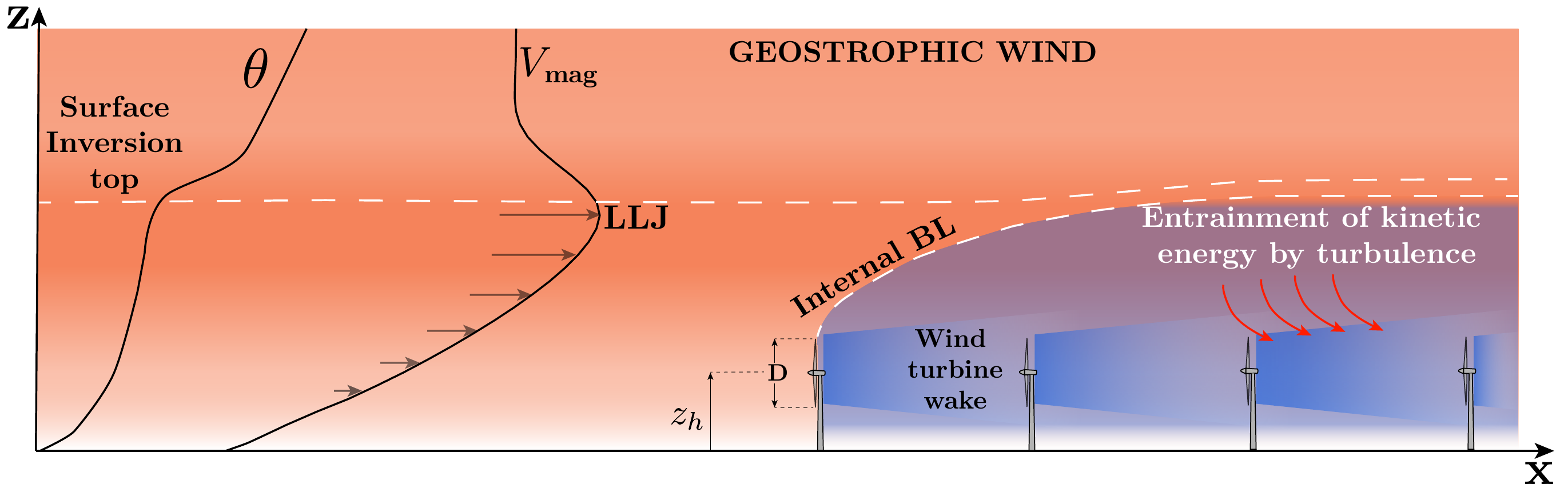}
 \end{subfigure}
 \caption{Sketch of the essential flow phenomena in wind farms in a SBL, including wakes and their superposition, the entrainment of {\color{black}energy} from above, and the development of the IBL. On the left, the typical temperature and velocity profiles, which reveal the LLJ and the top of the surface inversion, are sketched.}
 \label{fig1}
\end{figure}

The atmospheric boundary layer is dynamic and undergoes continuous transitions during the day due to changes in, for example, the surface heat flux and the geostrophic wind. The boundary layer is stably stratified in evenings due to cooling at the ground, and the wind in the residual layer decouples from the surface friction. Consequently, the balance between the Coriolis, frictional, and pressure forces is disturbed, and the flow in the residual layer accelerates. The acceleration produces a super-geostrophic jet at the top of the nocturnal stable boundary layer (SBL) at heights between 50 and 1000 meters \citep{sme96}. This super-geostrophic wind is known as a low-level jet (LLJ), and it generally forms due to the frictional decoupling combined with inertial oscillations \citep{bla57, tho77}. LLJs can also form due to large-scale baroclinicity or the pressure gradient due to cooling over sloped terrains \citep{mah99}. {\color{black} LLJs often form in nocturnal conditions when there is surface cooling \cite{baa09}.}
{\color{black}Mahrt (1998)} \cite{mah98} classified the nocturnal boundary layer into three stability regimes: 1) the weakly stable regime, characterized by continuous turbulence and a small downward heat flux, which is limited by the temperature fluctuations, 2) transition stability regime, where the quantities change rapidly with the increasing stability and the downward heat flux reaches a maximum, and 3) the very stable regime where the downward heat flux is small, limited by the turbulent vertical fluctuations, which are suppressed by buoyancy. High shear and weak to moderate stability characterize LLJs of practical importance \citep{baa09, ban08}. The shear in the LLJ is strong enough to generate continuous turbulent flux, with maximum and minimum turbulent flux near the surface and top of the SBL, respectively \citep{mah98}.\\
 %\citep{kal19, baa09}
\indent LLJs are frequently observed in many parts of the world, with occurrences in the Western ghats of India \citep{pra11}, the Great Plains of the United States \citep{kel04, ban02, lun03} and the Baltic sea of Europe \citep{sme93}.
% removed some things here as it was repetitive.
{\color{black} In the North Sea region LLJs at heights between 50--200 meters are observed with a frequency of 7.56\% in summer and 6.61\% during spring \cite{dun18}. Wind resource relevant LLJs in the North Sea are generally observed under stably stratified conditions \cite{baa09}. Therefore, the relevance of studying the impact of LLJs on wind farm applications has been emphasized by van Kuik et al. \cite{kui16} in the long-term European Research Agenda and a recent review by \citeauthor{por20} \citep{por20}.} 
%LLJs, therefore, have a direct impact on wind energy-based power production \citep{sis78}.
It is a common practice in wind power assessment to use simple power-law velocity profiles. However, this neglects the effects of LLJs on power production and the estimated fatigue loads \citep{gut17}. {\color{black}For example, LLJs have been found to increase the capacity factors by over $60\%$ under nocturnal conditions \cite{wil15b}}. However, as modern wind turbines are reaching heights above $200$ meter due to which interactions with LLJs become unavoidable. Consequently, it is imperative to study the interaction between LLJs and wind farms.\\
\indent When a large number of wind turbines operate in a wind farm, the structure of the boundary layer changes due to the momentum extraction by the turbines. Both numerical simulations and wind tunnel experiments show the development of an internal boundary layer (IBL) at the entrance of the wind farm \citep{fra06, cha11c}. Further downwind, in the fully developed regime, all the momentum is derived from vertical entrainment \citep{cal10, cal10b}. Due to the simplicity, most wind farm and atmospheric boundary layer simulations in the past have focused on pressure-driven neutral boundary layers. The underlying assumption in such simulations is that the wind turbines reside in the inner regions of the atmospheric boundary layer, where the outer layer effects, such as the rotation of the Earth and thermal stratification, are negligible \citep{cal10, ste17}. However, the wake recovery and entrainment of fresh momentum from outside the IBL strongly depend on the atmospheric stratification \citep{abk15b}. Furthermore, the wind follows an Ekman spiral due to the Coriolis force, affecting the wind turbine wakes as well as the wind farm wake. In essence, neglecting the stratification and Coriolis forces is too simplistic when considering the performance of large wind farms.\\
\indent Large-eddy simulations (LES) have been used extensively to study turbulence in the atmosphere \citep{moe84, mas92}, and the interaction between the atmospheric boundary layer and wind farms \citep{ste17, ste14, mey10}. LES has been successfully used to simulate both convective and stable atmospheric boundary layers at both weak and moderate stratification \citep{mas89, nie93, mas90, sai00, kos00}. Numerical simulations of weak and moderately stratified SBL are easier because of the continuous turbulence and the absence of global intermittency \citep{mah14}. The simulations of highly stratified boundary layers are challenging due to the mesoscale motions, gravity waves, the unsteady nature of the boundary layers, and LLJs. Nocturnal LLJs under weak to moderate stratification can be studied with LES \citep{bea06, kos00}, {\color{black}while boundary layers at higher stratification, for which the turbulence is intermittent and not continuous, is challenging to simulate with LES. We consider moderately stratified boundary layers in this paper.}\\
\indent Recently, the impact of the `capping' inversion on the power production of `infinitely' wide wind farms in conventionally neutral boundary layers is reported in the literature \cite{all17}. It has been found that the IBL pushes the capping inversion upwards, which generates pressure perturbations that travel upwind as gravity waves and slow down the in front of the wind farm. Furthermore, recent studies of wind farms {\color{black} in a neutral-to-stable boundary layer transition show that in a steady-state SBL the LLJ impacts the power production \cite{all18}.} Also, measurements and LES studies of wind farms in a SBL \citep{all18, dor15, wit14} show that due to low turbulence intensity, wake recovery is reduced compared to the unstable and neutrally stratified boundary layer. Besides, the rotation of the Earth affects the power production through the Coriolis forces, which deflects the wind farm wake \citep{van17b}. For specific wind directions, it has been found that even the horizontal component of the Earth's rotation influences the turbulent fluxes in a wind farm \citep{how20}. Furthermore, the vertical wind veer in the Ekman spiral causes a skewed spatial structure of the turbine wake, which enhances the shear production of turbulent kinetic energy leading to larger flow entrainment and faster wake recovery \citep{abk16b}. It is a common practice in the wind energy community to use periodic boundary conditions in the spanwise direction, which results in `infinitely' wide wind farms \citep{cal10, ste14, all15, wu13}. However, in the presence of Coriolis force, which induces appreciable wind veer, this assumption might lead to under-prediction of turbine power production, which directly interact with the LLJ. \\
\indent Previously, wind turbine and LLJ interactions have been studied by {\color{black} Lu and Port\'e-Agel (2011) \cite{lu11}}, who performed LES of the flow over a turbine in a doubly periodic domain (an `infinite' wind turbine array) with actuator line modeling, and they report non-axisymmetric turbine wakes and LLJ elimination due to energy extraction by the turbines. {\color{black}A similar study on the interaction between a single turbine and LLJ reports slower wake recovery at higher stratification and LLJ elimination \cite{bha15}. Furthermore, the LLJ weakening due to wind turbine energy extraction is also reported in diurnal cycle wind farm simulations \cite{abk16, sha17d}}. {\color{black}A similar phenomenon has been observed in the mesoscale weather model simulations of an infinite wind farm \cite{fit13b}}. Recently, Na et al. \cite{na18} performed LES of a small wind farm with 12 turbines arranged in three columns and four rows with an LLJ above it in a spanwise periodic domain. They report faster wake recovery due to the enhanced vertical kinetic energy flux created by the LLJ.\\ 
\indent The studies mentioned above have not addressed the effect of changing LLJ height on wind farm power production and do not provide a complete picture of the interaction between LLJs and wind farms as they consider spanwise `infinite' or very small wind farms. However, it is necessary to understand the coupling between stable stratification, flow-adjustment, and LLJ height on wind farm power production better. Figure \ref{fig1} shows the essential flow physics of the stable boundary layer wind farm interaction such as the IBL growth, turbine wake recovery, surface inversion, and the entrainment of momentum from above by turbulence.

{\color{black}In this work, we study the power production of a finite wind farm under stable stratification.} The objective of the study is two-fold, first to understand the effect of LLJ height and stable stratification on the power production of a wind farm, and second to study the effect of stable stratification on the flow adjustment in and around a `finite' wind farm. We study the wind farm - LLJ interaction by systematically reducing the surface cooling rate which produces LLJs of different heights. 

The remainder of the paper is structured as follows. In \autoref{sec2} the numerical method is explained. In \autoref{sec3} important boundary layer properties, the IBL growth above the wind farm, and the flow adjustment around the wind farm are discussed. In \autoref{sec4}, we carry out an analysis of the different flow phenomena by performing an energy budget analysis. Furthermore, in \autoref{sec5}, the effect of the wind veer is discussed, followed by the conclusions in \autoref{sec6}.

%%%%%%%%%%%%%%%%%%%%%%%%%%%%%%%%%%%%%%%%%%%%%%%%%%%%%%%%%%%%%%%%%%%%%%%%%%%%%%%%%%%%%%
\section{Large-eddy simulations}\label{sec2}
%%%%%%%%%%%%%%%%%%%%%%%%%%%%%%%%%%%%%%%%%%%%%%%%%%%%%%%%%%%%%%%%%%%%%%%%%%%%%%%%%%%%%%

In LES, the flow features larger than the filter size are fully resolved, while the sub-filter size eddies are modeled. {\color{black} Our code based on the one developed by \citeauthor{alb99} \citep{alb99, cal10}, which has been successfully updated with the dynamic, Lagrangian averaged scale-dependent model \cite{bou05}, actuator disk model for turbine modeling \citeauthor{cal10} \cite{cal10}, concurrent precursor method \citeauthor{ste14} \cite{ste14}, and thermal stratification \cite{nag19}.} This updated code has been validated for neutral and SBLs as well as the flow through wind farms \citep{ste16, zha19, nag19}. The governing equations and numerical method are discussed in \autoref{sec2.1}, and the boundary layer initialization is explained in \autoref{sec2.2}, followed by the wind farm setup in \autoref{sec2.3}. 

%%%%%%%%%%%%%%%%%%%%%%%%%%%%%%%%%%%%%%%%%%%%%%%%%%%%%%%%%%%%%%%%%%%%%%%%%%%%%%%%%%%%%%
\subsection{Governing equations and numerical method} \label{sec2.1} 
%%%%%%%%%%%%%%%%%%%%%%%%%%%%%%%%%%%%%%%%%%%%%%%%%%%%%%%%%%%%%%%%%%%%%%%%%%%%%%%%%%%%%%
The LES code we use integrates the filtered Navier-Stokes equations written for a wall-bounded turbulent flow \citep{alb96} and employs the Boussinesq approximation to model buoyancy. The governing equations are:
\begin{align}
 \partial_{\mathit{i}} \widetilde{u}_\mathit{i}&=0,\label{eqn2.1}\\
 \begin{split}
 \partial_{\mathit{t}}\widetilde{u}_\mathit{i} + \partial_\mathit{j}\left(\widetilde{u}_\mathit{i}\widetilde{u}_\mathit{j}\right)&=-\partial_{\mathit{i}}\widetilde{p}-\partial_\mathit{j}\tau_{\mathit{ij}} + g\beta(\widetilde{\theta}-\widetilde{\theta}_\mathit{0})\delta_{\mathit{i3}}+f_c(U_g-\widetilde{u})\delta_{i2}\\&\quad-f_c(V_g-\widetilde{v})\delta_{i1}+ \widetilde{f}_x\delta_{i1}+ \widetilde{f}_y\delta_{i2},
 \end{split}\label{eqn2.2}\\
 \partial_{\mathit{t}}\widetilde{\theta} + \widetilde{u}_\mathit{j}\partial_{\mathit{j}}\widetilde{\theta}&=-\partial_{\mathit{j}}q_\mathit{j},\label{eqn2.3}
\end{align}
where the tilde represents spatial filtering, $\widetilde{u}_\mathit{i}=\left(\widetilde{u},\widetilde{v},\widetilde{w}\right)$ and $\widetilde{\theta}$ are the filtered velocity and potential temperature, respectively, $g$ is the acceleration due to gravity, $\beta=1/\theta_\mathit{0}$ is the buoyancy parameter with respect to the reference potential temperature $\theta_\mathit{0}$, $\delta_{\mathit{ij}}$ is the Kronecker delta, $f_c$ is the Coriolis parameter. The boundary layer is driven by a mean pressure gradient, $p_\infty$, represented by the geostrophic wind with, $U_g=-\frac{1}{{\rho}f_c}\frac{\partial{p_{\infty}}}{\partial{y}}$ and $V_g=\frac{1}{{\rho}f_c}\frac{\partial{p_{\infty}}}{\partial{x}}$ as its components. $\widetilde{p}=\widetilde{p}^{*}/\rho+\sigma_{kk}/3$, is the modified pressure obtained by adding the trace of the sub-filter scale stress, $\sigma_{kk}/3$, to the kinematic pressure or pressure perturbation, $\widetilde{p}^{*}/\rho$, where $\rho$ is the density of the fluid. $\widetilde{f}_i=(\widetilde{f}_x,\widetilde{f}_y,0)$ represents the turbine forces, which are modeled using a filtered actuator disk approach \citep{jim10, cal10, cal11}. The molecular viscosity is neglected as it is a high Reynolds number flow, which is a common practice in atmospheric boundary layer simulations. $\tau_{\mathit{ij}}=\widetilde{u_{\mathit{i}}u_{\mathit{j}}}-\widetilde{u}_\mathit{i}\widetilde{u}_\mathit{j}$ is the traceless part of the sub-filter scale stress tensor, and $q_\mathit{j}=\widetilde{u_\mathit{j}\theta}-\widetilde{u}_\mathit{j}\widetilde{\theta}$ is the sub-filter scale heat flux tensor. The sub-filter stresses and heat fluxes are modeled as, 
\begin{align}
 \tau_{\mathit{ij}}&=\widetilde{u_{\mathit{i}}u_{\mathit{j}}}-\widetilde{u}_\mathit{i}\widetilde{u}_\mathit{j}=-2\nu_{T}\widetilde{S}_{ij}=-2(C_s\Delta)^2|\widetilde{S}|\widetilde{S}_{ij},\label{eqn2.4}\\
 q_\mathit{j}&=\widetilde{u_\mathit{j}\theta}-\widetilde{u}_\mathit{j}\widetilde{\theta}=-\nu_\theta\partial_j\widetilde{\theta}=-(D_s\Delta)^2|\widetilde{S}|\partial_j\widetilde{\theta},\label{eqn2.5}
\end{align}
where $\widetilde{S}_{ij}=\frac{1}{2}\left(\partial_j{\widetilde{u}_i} + \partial_i{\widetilde{u}_j}\right)$ is the filtered strain rate tensor, $\nu_T$ is the eddy viscosity, $C_s$ is the Smagorinsky coefficient for the sub-filter stresses, $\Delta$ is the filter size, $\nu_\theta$ is the eddy heat diffusivity, $D_s$ is the Smagorinsky coefficient for the sub-filter scale heat flux, and $|\widetilde{S}| = \sqrt{2\widetilde{S}_{ij}\widetilde{S}_{ij}}$. {\color{black} We use a tuning-free, scale-dependent model based on Lagrangian averaging of the coefficients \citep{bou05, sto06, sto08} to calculate the Smagorinsky coefficient dynamically. The error in the calculation of the Smagorinsky coefficients is minimized over fluid pathlines preserving the local fluctuations of the coefficients. The model uses a test-filter to calculate the coefficients dynamically and a second test-filter is used to overcome the limitation of scale-invariance \cite{bou05}, which makes the model particularly suitable for inhomogeneous flows, such as the flow through a wind farm or over complex terrain.}\\
%\\
{\color{black} \indent We employ a well-validated actuator disk model approach as is common in the simulation of large wind farms \cite{jim07, jim08,cal10, ste14, ste16, zha19, nag19}. The streamwise and spanwise compoments of the turbine force included in the momentum equation are given by $\tilde{f}_x=F_t\cos{\phi}$, and $\tilde{f}_y=F_t\sin{\phi}$, where $\phi$ is the angle the actuator disk makes with the x-axis, and $F_t$ is the turbine force modeled as, 
\begin{equation}
 F_t = -\frac{1}{2}\rho{C_T}{U^2_\infty}\frac{\pi}{4}D^2,\label{eqn:force}
\end{equation} 
where $C_T$ is the thrust coefficient and $U_\infty$ is the upstream undisturbed reference velocity. Equation \eqref{eqn:force} is only applicable for isolated turbines \cite{jim07, jim08} since the upstream velocity $U_\infty$ cannot be readily specified in wind farm simulations. Consequently, it is common practice \cite{cal10, cal11} to use actuator disk theory to relate $U_\infty$ with the rotor disk velocity $U_d$,
\begin{equation}
U_\infty= \frac{U_d}{\left(1-a\right)}\label{eqn:uinfty}
\end{equation} 
where $a$ is the axial induction factor. The total thrust force exerted by the turbines obtained by substituting equation \ \eqref{eqn:uinfty} in equation \eqref{eqn:force}: 
\begin{equation}
 F_t = -\frac{1}{2}\rho{C'_T}{\left<\overline{u}^T\right>^2_d}\frac{\pi}{4}D^2,\label{eqn:force2}
\end{equation}
where subscript d represents the averaging over the turbine disk, superscript $T$ represents averaging of the disk averaged velocity over time, and $C'_T=C_T/(1-a)^2=1.33$. For a detailed description and validation of the employed actuator disk model, we refer to Refs.\ \cite{cal10, cal11,ste18}. It is worth mentioning that the actuator disk model cannot capture the vortex structures near the turbine due to the absence of the turbine blades \cite{sor11, tro10, ste17}, which can be captured using an actuator line model. However, it is well established that the actuator disk model can accurately capture the wake dynamics, starting from 1 to 2 diameters downwind of the turbine \cite{ste17, ste18}. Therefore, the actuator disk model is commonly used to study the large scale flow phenomena in wind farms.}\\
%%%
\indent We use a pseudo-spectral method to calculate the partial derivatives in the streamwise and spanwise directions. The vertical direction is treated with a second-order central difference method. The solution is advanced in time by a second-order accurate Adams-Bashforth scheme. The aliasing errors resulting from the folding back of high wavenumber energy to the resolved scales due to the calculation of non-linear terms in physical space is prevented by using a $3/2$ anti-aliasing method \citep{can88}. For pointwise energy conservation, the convective term in the equation \eqref{eqn2.2} is written in the rotational form \citep{fer02}. More information about the numerical method can be found in \cite{alb96}. The computational domain is discretized uniformly with $n_x$, $n_y$, and $n_z$ points in the streamwise, spanwise, and vertical directions, respectively. Therefore, the corresponding grid sizes are $\Delta_{x} = L_x/n_x$, $\Delta_{y} = L_y/n_y$, and $\Delta_{z}=L_z/n_z$, where $L_x$, $L_y$, and $L_z$ are the dimensions of the computational domain. The computational grid is staggered in the vertical direction with the first grid point for $\widetilde{u}, \widetilde{v}$, and $\widetilde{\theta}$ located at a distance $\Delta_z/2$ above the ground. The computational plane for the vertical velocity, $\widetilde{w}$, is located at the ground. No-slip and free-slip boundary conditions with zero vertical velocity, $\widetilde{w}=0$, are used at the top and bottom boundaries, respectively. In wall-modeled LES of atmospheric boundary layers, the first grid point generally lies in the surface layer and the Monin-Obukhov similarity theory \citep{moe84} can be used to model the instantaneous shear stress $\tau_{i3|w}$ and buoyancy flux $q_{*}$ at the wall as follows: 
\begin{align}
\tau_{i3|w}=-{u_{*}^2}\frac{\widetilde{u}_i}{\widetilde{u}_r}=-\Bigg(\frac{\widetilde{u}_r\kappa}{\text{ln}(z/z_o)-\psi_{M}}\Bigg)^2\frac{\widetilde{u}_i}{\widetilde{u}_r},\label{eqn2.6}
\end{align}
and
\begin{align} 
q_{*}&=\frac{u_{*}\kappa(\theta_s-\widetilde{\theta})}{\text{ln}(z/z_o)-\psi_{H}},\label{eqn2.7}
\end{align}
where $\widetilde{u}_i$ and $\widetilde{\theta}$ represents the filtered velocities and potential temperature at the first grid point respectively, $u_*$ is the frictional velocity, $z_o$ is the roughness length, $\kappa$ is the von K\'arm\'an constant, $\widetilde{u}_r=\sqrt{\widetilde{u}^2 + \widetilde{v}^2}$ is filtered velocity magnitude at the first grid level, and $\theta_s$ is the filtered potential temperature at the surface. $\psi_M$ and $\psi_H$ are the stability corrections for momentum and heat flux, respectively. For SBLs, we use the stability correction suggested by \cite{bea06}, i.e.\ $\psi_{M}= -4.8z/L$ and $\psi_{H}= -7.8z/L$, where $L=-({u_*}^3\theta_{0})/({\kappa}gq_{*})$ is the surface Obukhov length. {\color{black} The wall-model is implemented as explained in Bou-Zeid et al.\ \cite{bou05}, i.e.\ the wall-layer fluxes and stresses are calculated using the filtered velocities at the first grid point above the ground.}
%
%%%%% Write about the code validation and stuff%%%%%%%%%
%
\subsection{Boundary layer initialization}\label{sec2.2}
Simulating strongly stratified boundary layers with LES is complicated due to the presence of globally intermittent turbulence \citep{mah14}. In the present work, we consider a moderately stable boundary layer ($z_i/L\approx2$, where $z_i$ is the boundary layer height). The boundary layer represents a typical quasi-equilibrium moderately SBL with a pronounced LLJ similar to those observed over polar regions and equilibrium night-time conditions over land in mid-latitudes. The case is well-documented under the global energy and water cycle experiment atmospheric boundary layer study (GABLS$-$1) initiative, and LES inter-comparison studies \citep{kos00, bea06}. The initial potential temperature profile has a mixed layer (with constant potential temperature 265 $\text{K}$) up to 100 $\text{m}$ with an overlying free atmospheric stratification of strength $0.01$ $\text{K}\text{m}^{-1}$. The reference potential temperature and roughness length are set to $263.5$ $\text{K}$ and $0.1$ $\text{m}$, respectively, and a constant surface cooling is applied. The boundary layer is driven by the geostrophic wind with the horizontal components $G=\left(U_g,V_g\right)=\left(8.0, 0.0\right)$ $\text{m}\text{s}^{-1}$. The Coriolis parameter is set to $f_c=1.39\times10^{-4}$ $\text{s}^{-1}$ (corresponding to latitude $\ang{73}\text{N}$). The initial wind profile is set equal to the geostrophic wind. Uniformly distributed random perturbations with an amplitude of $3\%$ of the geostrophic wind are added to velocities below a height of 50 $\text{m}$ to spin up turbulence. Similarly, uniformly distributed random perturbations with a magnitude of $0.1$ $\text{K}$ are added to the initial temperature profile. Detailed information about the SBL can be found in \citeauthor{bea06} \cite{bea06}.\\
\indent The boundary layer reaches a quasi-steady state at the end of 8$^\text{th}$ hour. The quasi-steady-state is said to have been reached when the temperature profile changes at a constant rate while the velocity and other turbulent quantities have reached a steady-state \citep{kos00}. Our code has been validated for the GABLS-1 boundary layer with a cooling rate of $0.25$ $\text{K}\cdot\text{hour}^{-1}$. In agreement with the previous study by Stoll and Port\'e-Agel (2008) \cite{sto08}, we found that the Lagrangian averaged scale-dependent (LASD) model produces better results when compared with the Smagorinsky model. We performed five simulations with surface cooling rates $C_r=[0.0, 0.125, 0.25, 0.375, 0.5]$ $\text{K}\cdot\text{hour}^{-1}$. Details about the different cases are documented in table \ref{table1}. It is also worth mentioning here that the potential temperature profile and the jet heights obtained with our simulations are similar to the ones observed in the North Sea \cite{baa09} and also the Dutch offshore wind Atlas (DOWA) simulation campaign.\\
\indent {\color{black} In addition to the stable cases, a reference case at truly neutral stratification, similar to Stevens et al.\ (2014) \cite{ste14} is also performed. Coriolis forces and thermal stratification are neglected for this case, and the boundary layer is driven by a mean pressure gradient $1/\rho({\nabla{p}})=-u_*^2$, where $u_*$ represents the friction velocity of near-neutral stratification case with $C_r=0.0$ $\text{K}\cdot\text{hour}^{-1}$, i.e.\ SBL--1. This is the truly neutral boundary layer (TNBL), which has a logarithmic velocity profile without LLJ.}
\subsection{Wind farm setup}\label{sec2.3} 
\begin{figure}
 \centering
 \begin{subfigure}[h]{0.99\textwidth}
 \includegraphics[width=\linewidth]{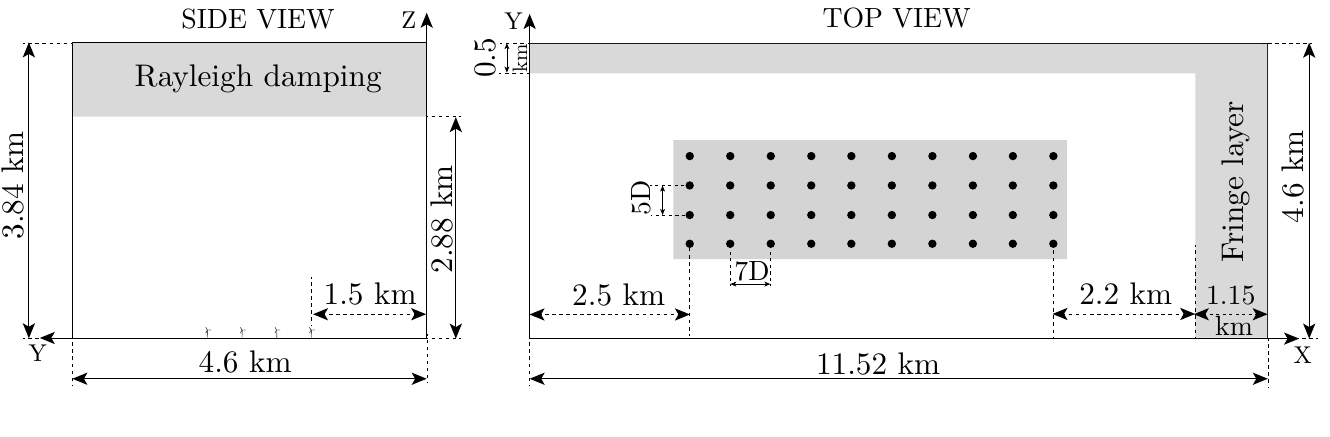} 
 \end{subfigure}
 \caption{Schematic of the computational domain, showing the wind farm layout, the extent of Rayleigh damping layer, and the fringe layers. Black circles indicate the positions of the wind turbines. The statistics are sampled from the shaded region of size $70D\times{20D}\times{D}$, which is centered around the wind farm.}
 \label{fig2}
 \vspace{-4mm}
\end{figure}

We consider a large wind farm with 40 wind turbines. The turbines are distributed in an array of $4$ columns and $10$ rows. The turbine diameter is $D=90$ $\text{m}$ and the hub height is $z_h=90$ $\text{m}$. The turbines are separated by a distance of $s_x=7D$ and $s_y=5D$ in the streamwise and spanwise directions, respectively. The computational domain is $11.52$ $\text{km}$ $\times$ $4.6$ $\text{km}$ $\times$ $3.84$ $\text{km}$. The details of the computational domain and wind farm layout are given in figure \ref{fig2}. According to the Monin-Obukhov similarity theory, the first grid point above the ground should be in the inertial sublayer. For the GABLS-1 case, cautioning against using very high resolution near the ground, which violates the similarity theory, Basu and Lacser (2017) \cite{bas17} suggest using $z_1\geq50z_o$, where $z_1$ the height of the first grid point above the ground. Accordingly, we fix the vertical grid resolution to be $5$ $\text{m}$. We use a horizontal resolution of 9 m in the streamwise and spanwise direction to ensure that the important flow scales are properly resolved. As a result, the domain is discretized by $1280\times512\times768$ grid points in the streamwise, spanwise, and vertical directions, respectively. The computational domain has approximately 500 million grid points.\\
\indent We use a large vertical extent and a Rayleigh damping layer \citep{kle78} with a strength of $0.016$ $\text{s}^{-1}$ in the top $25$\% of the domain to reduce the effects of gravity waves. We find that this damps out most of the generated gravity waves. To ensure that the streamwise fringe layer does not affect the turbulence statistics, we performed a simulation in a bigger domain of size $17.28$ $\text{km}$ $\times$ $4.6$ $\text{km}$ $\times$ $3.84$ $\text{km}$ with a resolution of $18$ $\text{m}$ $\times$ $18$ $\text{m}$ $\times$ $10$ $\text{m}$ and compared it with the results of the smaller domain. The streamwise fringe layer in the bigger domain is $75D$ downwind of the wind farm. We found that the streamwise domain size does not affect the turbulence statistics relevant to the study, which confirms that the used domain size is sufficient for the purposes of this study.\\
\indent To obtain realistic inflow conditions, we employ the concurrent precursor technique \citep{ste14b}. In this technique, simulations are run in two domains concurrently. We perform the atmospheric boundary layer simulations without wind turbines to generate inflow conditions in a precursor domain. Then the quantities from the precursor domain are used as the inlet conditions for the wind farm domain. The forcing is done in the wind farm domain by gradually blending the velocities in the fringe layer. An Ekman spiral, which induces considerable spanwise flow, is formed due to the action of the Coriolis forces. Therefore, we use fringe layers in both the streamwise and spanwise direction to eliminate the effects of the periodic boundaries. We fix the fringe layer length to be $10\%$ of the computational domain in the streamwise and spanwise directions, respectively.\\
\indent The equilibrium wind angle under geostrophic forcing depends on the stability conditions, which results in a different geometric pattern of the turbines and complicates the analysis. To ensure the same farm layout in all simulations, we use a proportional-integral (PI) controller \citep{ses14}, similar to the one used by \cite{all15}, to rotate the incoming flow such that the planar averaged wind angle at hub height is always zero. Even then, local changes in the wind angle upwind of a turbine can result in turbine yaw misalignment, which results in sub-optimal energy production. Each turbine in the simulations has an individual yaw-angle controller, which reorients the turbines perpendicular to the incoming wind direction measured $1D$ upwind of each turbine, to prevent yaw misalignment.

%%%%%%%%%%%%%%%%%%%%%%%%%%%%%%%%%%%%%%%%%%%%%%%%%%%%%%%%%%%%%%%%%%%%%%%%%%%%%%%%%%%%%%
\section{LES of a finite wind farm}\label{sec3}
%%%%%%%%%%%%%%%%%%%%%%%%%%%%%%%%%%%%%%%%%%%%%%%%%%%%%%%%%%%%%%%%%%%%%%%%%%%%%%%%%%%%%%
All the simulations are carried out in two stages. In the initial or the spin-up stage, only the SBL in the precursor domain is considered. The SBL reaches a quasi-steady state at the end of the $8^\text{th}$ hour. In the second stage, the turbines are introduced in the main domain, and the simulation in both domains is continued concurrently for one more hour in which the transient effects of the turbine startup subside. Finally, both simulations are run for one more hour, and the statistics are collected in the last hour, i.e.\ the $10^{\text{th}}$ hour. {\color{black} Flow statistics were collected as ten-minute samples to quantify the uncertainty in the mean values. The standard deviation of the six samples over the mean is within $5\%$.} Basic boundary layer characteristics are presented in \autoref{sec3.1}, and the development of the IBL over the wind farm and the flow-adjustment are discussed in \autoref{sec3.2}. 
\begin{table}
 \begin{center}
\def~{\hphantom{0}}
 \caption{{\color{black} Details of the LES}. The columns from left to right indicate the case name, the surface cooling rate $C_r$, the boundary layer height $z_i$, the jet height $z_\text{jet}$, and the inversion height $z_c$. $u_\text{jet}$ is the velocity at jet height. $\mathrm{TI}=\sigma_u/u_\text{mag}$ is the turbulence intensity at hub height. All heights are normalized with the hub height.}
 \begin{tabular}{lcccccccccc}
 \hline
 Case & $C_r$ [$\text{K}\cdot\text{h}^{-1}$] & $z_i/z_h$ & $z_\text{jet}/z_h$ & $z_c/z_h$ &$u_*/G$ & $u_\text{jet}/G$ & $z_i/L$ & $TI_\mathsf{z_h}\%$ \\[3pt]
 \hline
 TNBL & -- & -- & -- & -- & 0.0395 & -- & -- & 10.70\\
 \hline
 SBL--1 & 0.000 & 2.839 & 2.670 & 3.023 & 0.0395 & 1.109 & 0.350 & 5.62\\
 \hline
 SBL--2 & 0.125 & 2.313 & 2.169 & 2.506 & 0.0348 & 1.157 & 1.103 & 4.29\\
 \hline
 SBL--3 & 0.250 & 1.903 & 1.836 & 2.114 & 0.0316 & 1.180 & 1.713 & 3.18\\
 \hline
 SBL--4 & 0.375 & 1.668 & 1.557 & 1.840 & 0.0296 & 1.187 & 2.274 & 2.40\\
 \hline
 SBL--5 & 0.500 & 1.551 & 1.446 & 1.639 & 0.0285 & 1.189 & 2.859 & 1.95\\
 \hline
 \end{tabular}
\label{table1}
\end{center}
\end{table}

%%%%%%%%%%%%%%%%%%%%%%%%%%%%%%%%%%%%%%%%%%%%%%%%%%%%%%%%%%%%%%%%%%%%%%%%%%%%%%%%%%%%%%
\subsection{Boundary layer properties}\label{sec3.1}
%%%%%%%%%%%%%%%%%%%%%%%%%%%%%%%%%%%%%%%%%%%%%%%%%%%%%%%%%%%%%%%%%%%%%%%%%%%%%%%%%%%%%%

\begin{figure}
 \begin{subfigure}[ht!]{0.75\textwidth}
 \includegraphics[width=\linewidth]{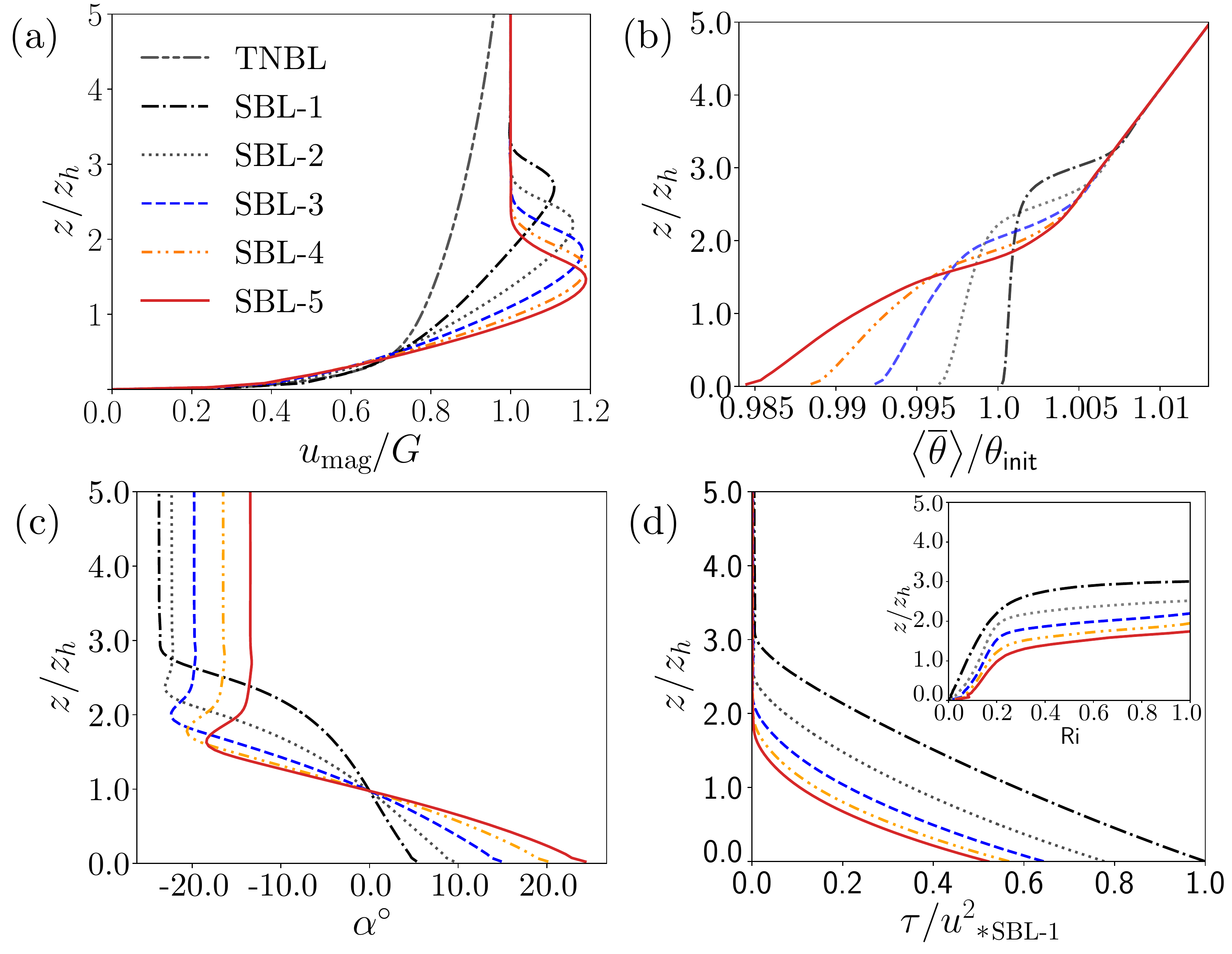}
 \end{subfigure}
 \caption{(a) Horizontal velocity magnitude, (b) potential temperature, (c) wind angle, and (d) {\color{black} vertical momentum flux} as a function of height for the different cases, see table \ref{table1} for details. The inset in panel (d) shows the variation of the gradient Richardson number with height.}
 \label{fig3}
\end{figure}

An overview of the surface forcings and the basic boundary layer properties such as the boundary layer height, friction velocity, jet velocity, and the stability parameter $z_i/L$ in the precursor domain are presented in table \ref{table1}. We determine the boundary layer height by the method used by \citeauthor{kos00} \cite{kos00} and \citeauthor{bea06} \cite{bea06}. The boundary layer height $z_i$ is defined as the height where the mean stress is $5\%$ of its surface value ($z_{0.05}$) followed by a linear extrapolation, i.e.\ $z_i = z_{0.05}/0.95$. At higher cooling rates, the friction velocity decreases, which indicates that there is reduced turbulence in the boundary layer. Furthermore, the boundary layer becomes shallower, i.e.\ $z_i$ reduces. 

Figure \ref{fig3}(a) presents the planar averaged horizontal wind magnitude $u_\text{mag}=\left<\sqrt{\overline{{u}}^2 + \overline{{v}}^2}\right>$, where $\left<\right>$ represents planar averaging, the overbar represents time averaging, and the tilde representing filtering is dropped in the remainder of the paper for simplicity. The strength of the jet, which is defined as the ratio of wind magnitude of the jet to the geostrophic velocity, i.e.\ $u_\text{jet}/G$, increases as the cooling rate increases while the jet height $z_\text{jet}/z_h$ decreases. The jet plays an important role in sustaining continuous turbulence in the boundary layer \citep{ban08, mah98}. For stronger stratification cases SBL--4 and SBL--5 (see table \ref{table1}), the ratio of the jet height to the turbine hub height is approximately $1.5$, which means that the jet height is equal to the height of the top of the turbine blades. {\color{black} The logarithmic velocity profile from the reference TNBL case is also presented in Fig.\ \ref{fig3}(a). We use the friction velocity $u_*=0.316$ m/s  obtained from the SBL--1 simulation with near-neutral stratification to ensure that the surface fluxes of the SBL--1 and TNBL case match. We note that a similar value (0.306--0.315 m/s) for the conventionally neutral boundary layers has been obtained by Allaerts and Meyers (2017) \cite{all17}. We find that close to the surface, i.e.\ $z/z_h<0.5$, the velocity profiles of TNBL and SBL--1 are nearly the same. The figure shows that stable cases have stronger shear than the TNBL case due to the jet's presence. The absolute power production increases with atmospheric stratification, which is explained in detail in section \ref{sec4.2}.\\}
\indent Figure \ref{fig3}(b) shows the planar averaged potential temperature profile. The height of the inversion $z_c$ is defined as the height where the temperature gradient is maximum. For $z_c/z_h \leq 1.8$, we observe that the inversion height is approximately equal to the SBL height, such that the direct interaction with the IBL developed by the wind farm is possible. Figure \ref{fig3}(c) presents the wind angle variation $\alpha$ as a function of height for the different cases. For higher cooling rates, a wind veer as strong as $\ang{15}-\ang{20}$ is observed. {\color{black} This wind veer also affects power production, which is significant in the presence of the jet, and the phenomenon is explained in detail in section \ref{sec5}.} 

Based on $z_i/L$, {\color{black}Holtslag and Nieuwstadt (1986) \cite{hol86}} identified three SBLs regimes, namely 1) near-neutral regime ($0<z_i/L\leq1$) with weak stability characterized by continuous turbulence, 2) an intermediate regime ($1<z_i/L\leq10$) with moderate stability where the boundary layer follows z-less scaling with continuous turbulence, and 3) a highly stable intermittency regime ($z_i/L>10$) where the turbulence is weak and sporadic and therefore not continuous in time and space. In all the cases considered in the present study, $z_i/L<3$, indicating weak to moderate stability of the boundary layers. Under such conditions the boundary layer remains continuously turbulent, and the similarity theory applies to the surface layer. Furthermore, continuous turbulence is sustained by the high shear of the LLJs.

In addition to $z_i/L$ the effect of inversion, which takes into account the free atmospheric stratification, can also be characterized by the gradient Richardson number ($\text{Ri}$) calculated by the Brunt-V{\"a}is{\"a}l{\"a} frequency $N$ and mechanical shear $S$: 
\begin{equation}
 \text{Ri}(z) = \frac{N^2}{S^2}; \quad N^2 = \frac{g}{\theta_0}\frac{\partial\left<\overline{\theta}\right>}{\partial{z}}; \quad S^2 = \left[\left(\frac{\partial{\left<\overline{u}\right>}}{\partial{z}}\right)^2 + \left(\frac{\partial{\left<\overline{v}\right>}}{\partial{z}}\right)^2\right]. 
\end{equation}
Figure \ref{fig3}(d) shows the planar averaged vertical momentum flux in the precursor domain. The planar averaged vertical momentum flux defined as $\tau=\left<\sqrt{(\overline{u'w'})^2 + (\overline{v'w'})^2}\right>$, where $\overline{u'w'}=\left(\overline{{uw}} + \overline{\tau_{xz}}\right)-\overline{{u}}~\overline{{w}}$, and $\overline{v'w'}=\left(\overline{{vw}} + \overline{\tau_{yz}}\right)-\overline{{v}}~\overline{{w}}$. The fluxes are normalized with the surface flux of the SBL--1 case to show the reduction in the turbulent momentum flux at higher cooling rates. It is evident from figure \ref{fig3}(d) that the turbulence in the boundary layer reduces when the surface cooling rate is increased. The inset in figure \ref{fig3}(d) shows that the Richardson number $\text{Ri}$ increases monotonically with height for all the cases. At the top $10$ to $20\%$ of the boundary layer, the $\text{Ri}$ increases above the critical $\text{Ri}_\text{c}$ (based on the hydrodynamic instability theory \cite{ric20, tay31, mil86, gal07}). Zilitinkevich et al. \cite{zil08} classify the boundary layer into three regimes: 1) weakly stable regime at $\text{Ri}<0.1$, 2) a transitional regime at $0.1\leq\text{Ri}\leq1$ with strong turbulence at $\text{Ri}<<1$; and 3) weak turbulence regime at $\text{Ri}>1$, capable of transporting momentum but not heat. At higher cooling rates (cases SBL--4 and SBL--5), the $\text{Ri}$ number increases rapidly with height, limiting the turbulence to very low heights, which affects the IBL dynamics in the presence of a wind farm. This means, above the LLJ there is negligible turbulence and wake recovery will be affected at lower jet heights. It is worth mentioning here that the turbulence intensity is maximum for the reference TNBL case.\\
\indent To conclude, the initialization stage yields completely turbulent, quasi-steady boundary layer which serves as an realistic inflow condition for the wind farm.

%%%%%%%%%%%%%%%%%%%%%%%%%%%%%%%%%%%%%%%%%%%%%%%%%%%%%%%%%%%%%%%%%%%%%%%%%%%%%%%%%%%%%%
\subsection{Flow adjustment in and around the wind farm}\label{sec3.2}
%%%%%%%%%%%%%%%%%%%%%%%%%%%%%%%%%%%%%%%%%%%%%%%%%%%%%%%%%%%%%%%%%%%%%%%%%%%%%%%%%%%%%%

\begin{figure}
 \centering
 \begin{subfigure}[ht!]{0.75\textwidth}
 \includegraphics[width=\linewidth]{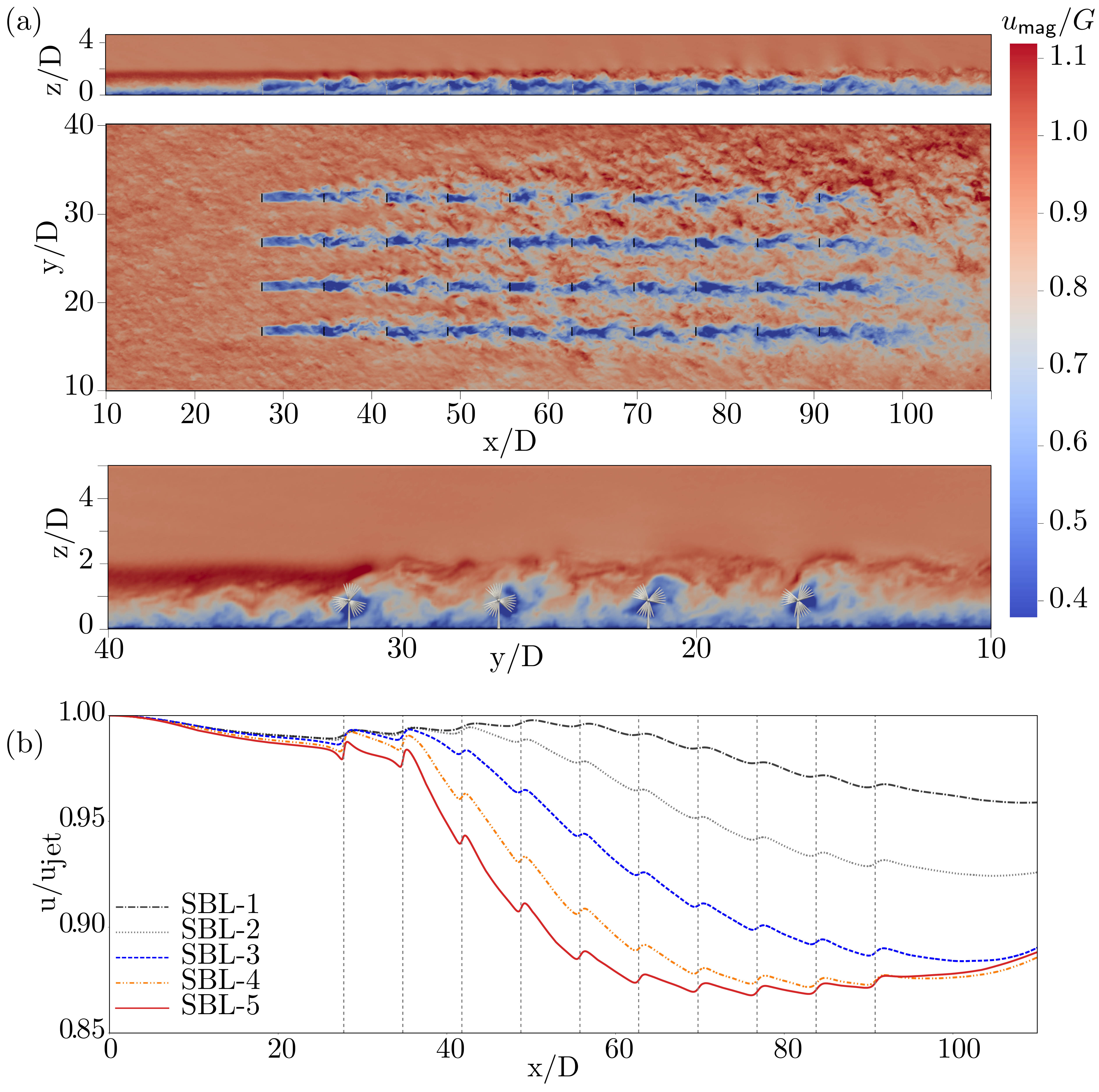}
 \end{subfigure}
 \caption{(a) Instantaneous velocity contours $u_\text{mag}/G$ for the SBL--3 case. Top panel: Side view of the wind farm in an $x-z$ plane through the middle of the 3rd turbine column from the bottom. Middle panel: Velocity contours at hub height ($x-y$ plane). Bottom panel: Front view of the wind farm in an $y-z$ plane passing through the fifth row. (b) Streamwise variation of the jet strength. Dashed vertical lines represent the turbine positions.}
 \label{fig4}
\end{figure}

\begin{figure}
 \centering
 \begin{subfigure}[ht!]{0.42\textwidth}
 \includegraphics[width=\linewidth]{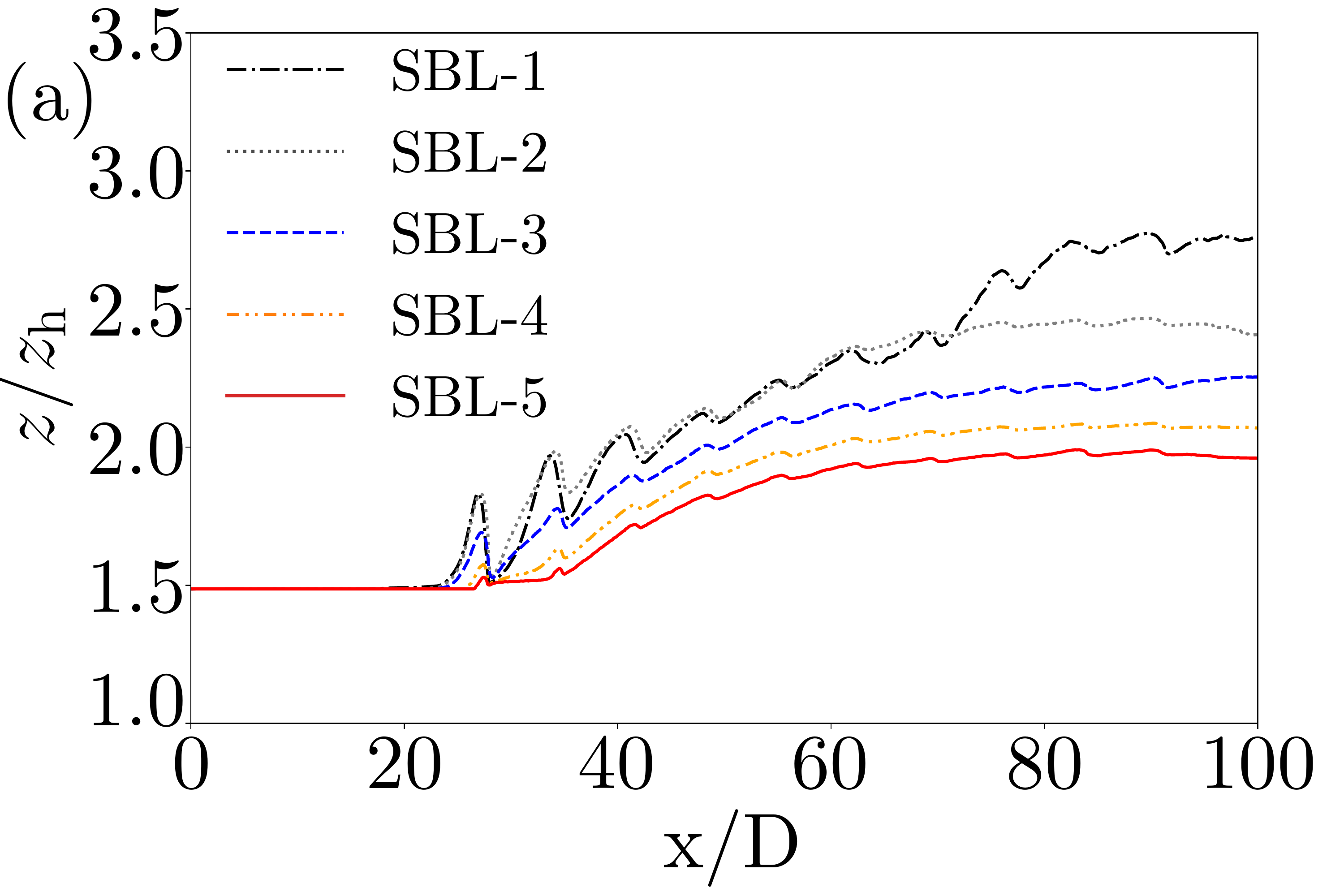}
 \end{subfigure}
 \begin{subfigure}[ht!]{0.42\textwidth}
 \includegraphics[width=\linewidth]{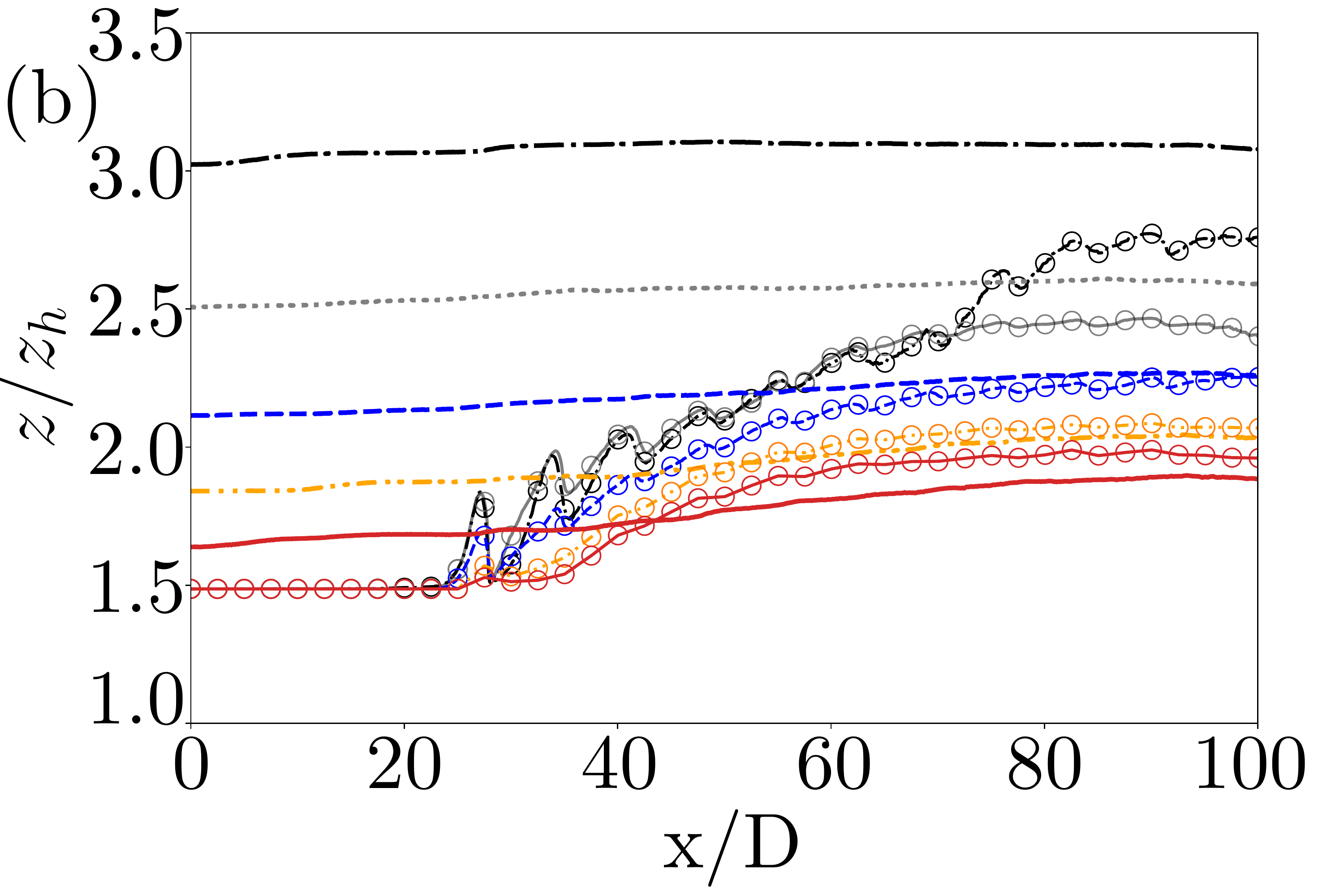}
 \end{subfigure}
 \caption{(a) The development of the IBL height with streamwise distance. (b) The lines indicate the top of the surface inversion, and the lines with markers the IBL height as in panel (a). Note that for SBL--4, and SBL--5 the IBL grows above the surface inversion.}
 \label{fig5}
\end{figure}

{\color{black} Instantaneous flow structures of the horizontal velocity for the SBL--3 case are shown in figure \ref{fig4}(a), and the streamwise variation of the jet velocity is presented in figure \ref{fig4}(b). The top panel of figure \ref{fig4}(a) illustrates the velocity contours in an $x-z$ plane (through the third column, note that only the lowest $z/D=5$ is shown). There is a steady wake behind the first turbine row, which does not interact much with the LLJ. Therefore, there is a negligible drop in the jet velocity behind the first turbine row; see the blue dashed curve in figure \ref{fig4}(b). Subsequently, the wake behind the second turbine row shows transverse wake meandering along with entrainment and the jet strength starts reducing. Wake meandering further adds to the background atmospheric turbulence \cite{mao18, lar08, fot19} and plays a significant role in entraining the jet's high-velocity fluid. The wakes interact with the LLJ in two ways: by wake meandering and by turbulent entrainment, both reduce the jet strength. This reduction in the jet velocity affects the power production, which will be explained in the energy budget analysis presented in section \ref{sec4.1}.\\
\indent The middle panel of figure \ref{fig4}(a) shows a horizontal snapshot of the flow at hub height ($x-y$ plane). We notice straight wakes behind the first turbine row and significant wake meandering in the lateral direction after the second turbine row. This shows that the onset of wake meandering is delayed when atmospheric stability is increased, which negatively affects the power production of the second row. The bottom panel in figure \ref{fig4}(a) shows a $y-z$ plane at a distance (1D) behind the sixth turbine row. This figure is interesting as it shows a significant spanwise flow of the fluid with the LLJ impinging on the turbine in the first column on the left. This happens due to the wind veer induced by the Coriolis forces. As a result, the turbines in the first column entrain the high-velocity jet, which increases the power production of that column. This effect is explained in more detail in section \autoref{sec5}. Another noteworthy point here is that the figure shows the importance of performing non-periodic, fully-finite simulations using a fringe layer in the spanwise direction.} {\color{black} In a spanwise `infinite' wind farm simulation, the turbine in the first column would be operating in the wake of the wind farm, due to which the turbine power production would be underpredicted.\\}
\indent The turbines extract energy from the incoming flow and thereby create a momentum deficit in the wake. The wakes start interacting with the boundary layer both in the lateral and vertical direction via turbulence, and the momentum deficit spreads in the boundary layer, which in turn entrains air towards the turbines. The region of momentum deficit gives rise to the IBL, above which the boundary layer is undisturbed by the dynamics near the surface. In contrast, inside the IBL, the flow structure changes downwind due to momentum extraction by the turbines. The growth of the IBL shows how the wind farm modifies the flow.
Furthermore, the height of the IBL is useful in the analytical modeling of wind farm power production \citep{men12}. There is no set rule for calculating the IBL height. For example, \cite{wu13} define it as the height where the time-averaged wake velocity is $99\%$ of the mean flow velocity at that height, {\color{black}Allaerts and Meyers (2017) \cite{all17}} define it as the height where the ratio of time-averaged horizontal velocity magnitude and the inflow velocity at the same height, taken in a plane 2 km upwind, reaches a threshold of $97\%$, and  {\color{black}Stevens (2014) \cite{ste14c}} defines it as the height where the vertical energy flux reaches the free stream value. We define the IBL as the height where the time-averaged horizontal velocity magnitude $u_\text{mag}$ is $97\%$ of the planar averaged inflow velocity at the same height. Besides, we fix the turbine top ($z_h+D/2$) as the minimum height of the IBL as the IBL grows over the turbine top. Figure \ref{fig5}(a) shows that the IBL height decreases when the surface cooling rate increases and grows with the downwind location in the wind farm. This is analogous to the growth of an IBL over a roughness change due to horizontal advection of air. Here, the presence of a wind farm is felt by the upwind flow as a roughness change, and due to the continuity constraint, the flow accelerates over the wind farm.\\ 
\indent In an atmospheric boundary layer, inversion represents a region where the potential temperature increases with height. In a SBL, the temperature increases with height from the ground and it is called surface inversion. The surface inversion top represents the height where the temperature gradient is maximum, above which the flow is non-turbulent. Due to the presence of the wind farm, the surface inversion top gets pushed up by the growing wind farm IBL. In figure \ref{fig5}(b), the top of the surface inversion $z_c$, defined as the height where the temperature gradient is maximum, is plotted along with the IBL for different cases. It is evident from the figure that the surface inversion top is pushed up due to the IBL. For the first two cases, the IBL stays below the inversion top. The displacement of the inversion top increases with the increased cooling rate, and for SBL--4 and SBL--5 the IBL grows above the surface inversion top. The wind above the surface inversion is non-turbulent in these cases, and the $\text{Ri}$ number of the flow is high at the top of the boundary layer. In these cases, the surface inversion top acts as a lid, limiting the growth of the IBL. Due to the continuity constraint, the wind goes around the wind farm. The space between the top of the turbines and the surface inversion top determines how much wind flows around the wind farm. The surface inversion top is at the height of $z_c/z_h\leq2.114$ for cases SBL--3, SBL--4, SBL--5, which is approximately $0.5D$ or less above the tip of the turbines. Consequently, the stabilizing effect of the surface inversion top restricts the growth of IBL in the vertical direction. Therefore, we see an appreciable amount of flow going around the wind farm. In essence, the so-called blockage due to the wind farm is the highest for SBL--5 and lowest for SBL--1.\\
\begin{figure}
 \centering
 \begin{subfigure}[ht!]{0.7\textwidth}
 \includegraphics[width=\linewidth]{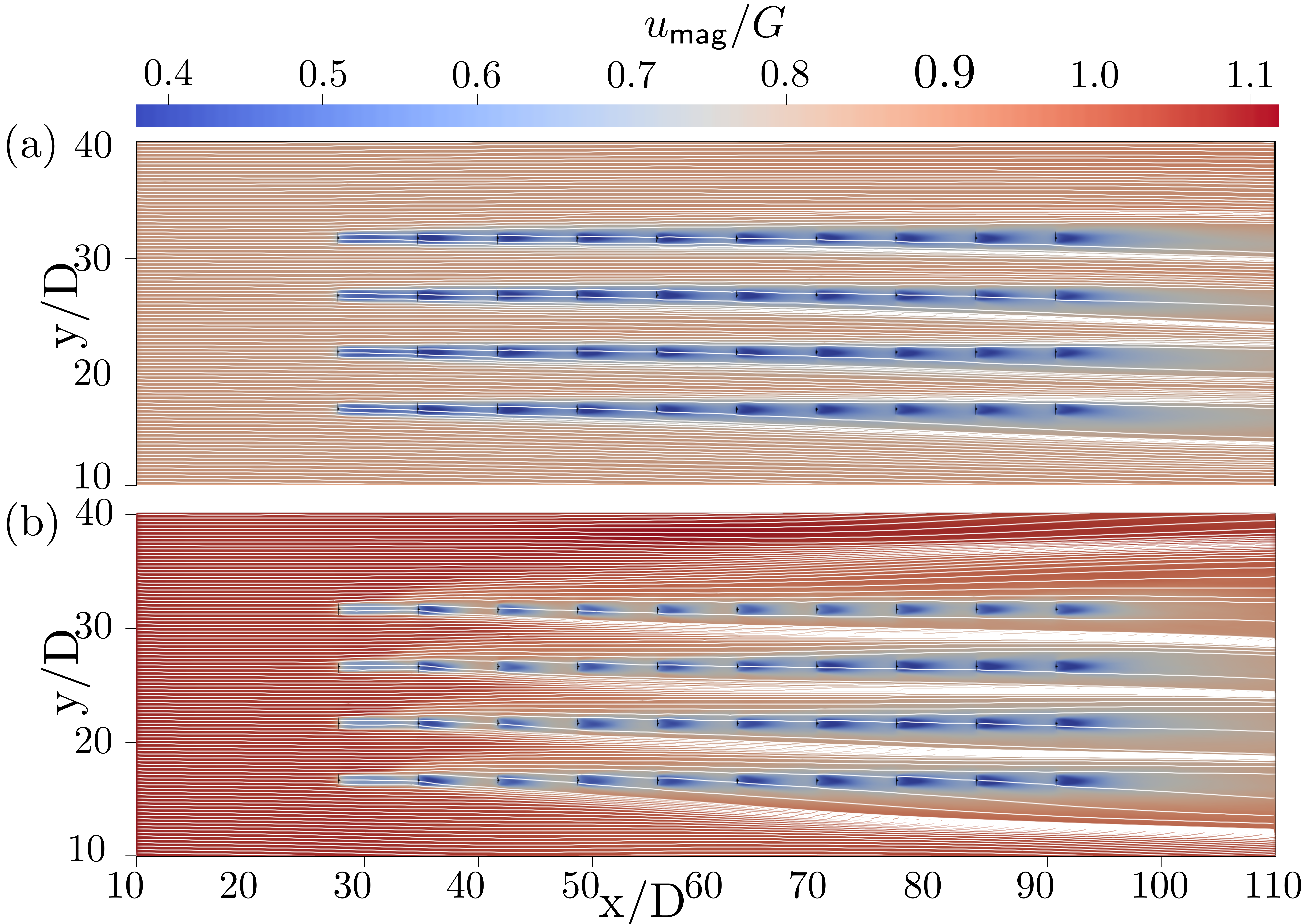}
 \end{subfigure}
 \centering
 \begin{subfigure}[ht!]{0.8\textwidth}
 \includegraphics[width=\linewidth]{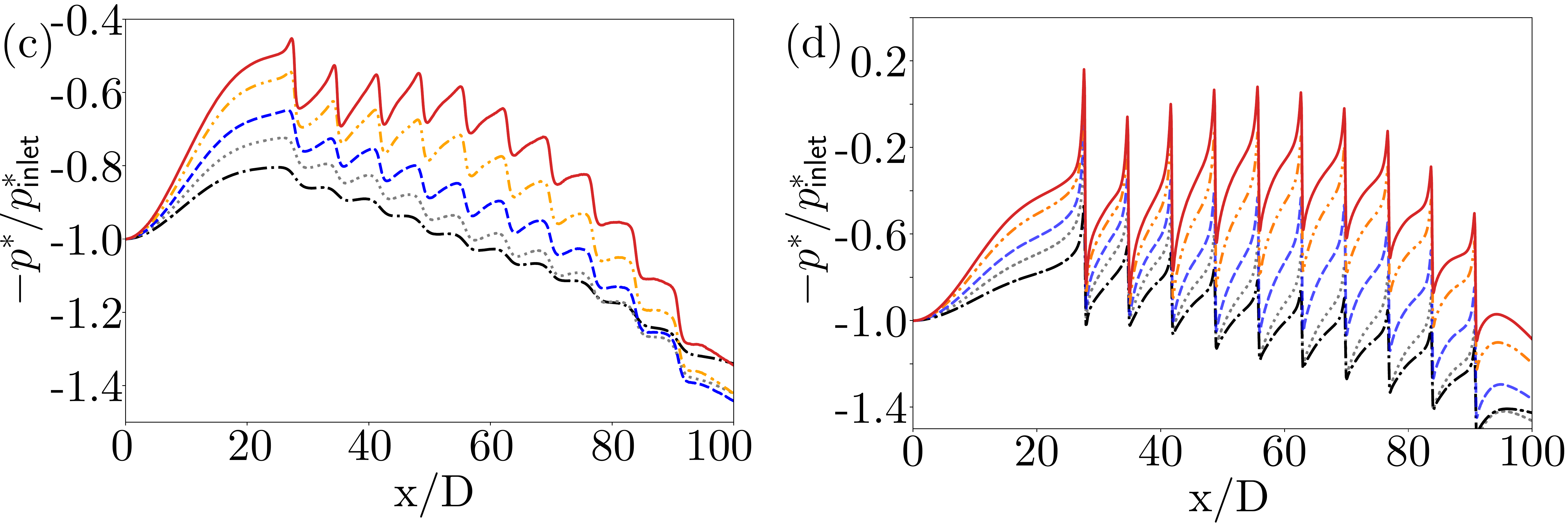}
 \end{subfigure}
 \caption{Streamlines at the hub height for the (a) SBL--1 and (b) SBL--5 cases. Note that for SBL--5 in which the IBL grows above the surface inversion, the streamlines indicate that there is very significant flow around the farm. (c) Pressure perturbation at the surface inversion top $z_c$ as a function of streamwise distance. The flow experiences maximum adverse pressure gradient for SBL--5. (d) The variation of pressure perturbation at hub height with the streamwise distance. In (c) and (d) $p^{*}_\text{inlet}$ is the pressure perturbation at the inlet.}
 \label{fig6} 
\end{figure}
\indent Figures \ref{fig6}(a,b) show the time-averaged streamlines at hub height for the cases SBL--1 and SBL--5. Figure \ref{fig6}(a) shows the streamlines for SBL--1; we see that the streamlines are nearly parallel and show marginal divergence. Figure \ref{fig6}(b) shows the streamlines for the SBL--5 case; we observe significant streamline divergence proving that the flow goes around the wind farm. {\color{black}Rominger and Nepf (2011) \cite{rom11}} observe that when a flow encounters the leading edge of a canopy, a part of the flow is diverted, and the remaining part advects through the porous canopy. As the turbines start extracting energy, the shear in the IBL reduces, causing an increase in the $\text{Ri}$ number in the IBL. The inset of figure \ref{fig3}(d) shows that the increase in $\text{Ri}$ with height is maximum for SBL--5. As the shear in the flow decreases due to the energy extraction by the turbines, the $\text{Ri}$ increases. With the increase in local $\text{Ri}$, the flow stability increases, and the fluid finds it challenging to go over the wind farm, and it takes the path of least flow resistance, i.e.\ around the wind farm. The effect is similar to the flow going around a three-dimensional obstacle like a mountain under highly stratified conditions \citep{hun80, bai79}.

Figures \ref{fig6}(c,d) show the pressure perturbation normalized by the inlet pressure at the top of the surface inversion $z_c$ and at the hub height for the different cases. For SBL--5, the pressure perturbation starts increasing in the entrance region of the wind farm when the IBL is at the same height as the surface inversion top. As this poses resistance to the developing IBL, the flow experiences an adverse pressure gradient; this makes it difficult for the flow to go through or over the wind farm, forcing it to go around.

%%%%%%%%%%%%%%%%%%%%%%%%%%%%%%%%%%%%%%%%%%%%%%%%%%%%%%%%%%%%%%%%%%%%%%%%%%%%%%%%%%%%%%
\section{Energy budget analysis}\label{sec4}
%%%%%%%%%%%%%%%%%%%%%%%%%%%%%%%%%%%%%%%%%%%%%%%%%%%%%%%%%%%%%%%%%%%%%%%%%%%%%%%%%%%%%%

In the boundary layer, the wind turbines extract energy from the flow and entrain fresh momentum from the upper layers of the atmosphere. An energy budget analysis is a convenient way to understand the diverse phenomena involved in the power production of a wind farm. {\color{black}We follow the budget analysis by Allaerts and Meyers \citep{all17} on wind farms in conventionally neutral boundary layers}. In \autoref{sec4.1}, a budget analysis of the total energy and its different components is presented, and the turbine power production is discussed in \autoref{sec4.2}.

\subsection{Entrainment, streamwise flow work}\label{sec4.1}
\begin{figure}
 \centering
 \begin{subfigure}[ht!]{0.8\textwidth}
 \includegraphics[width=\linewidth]{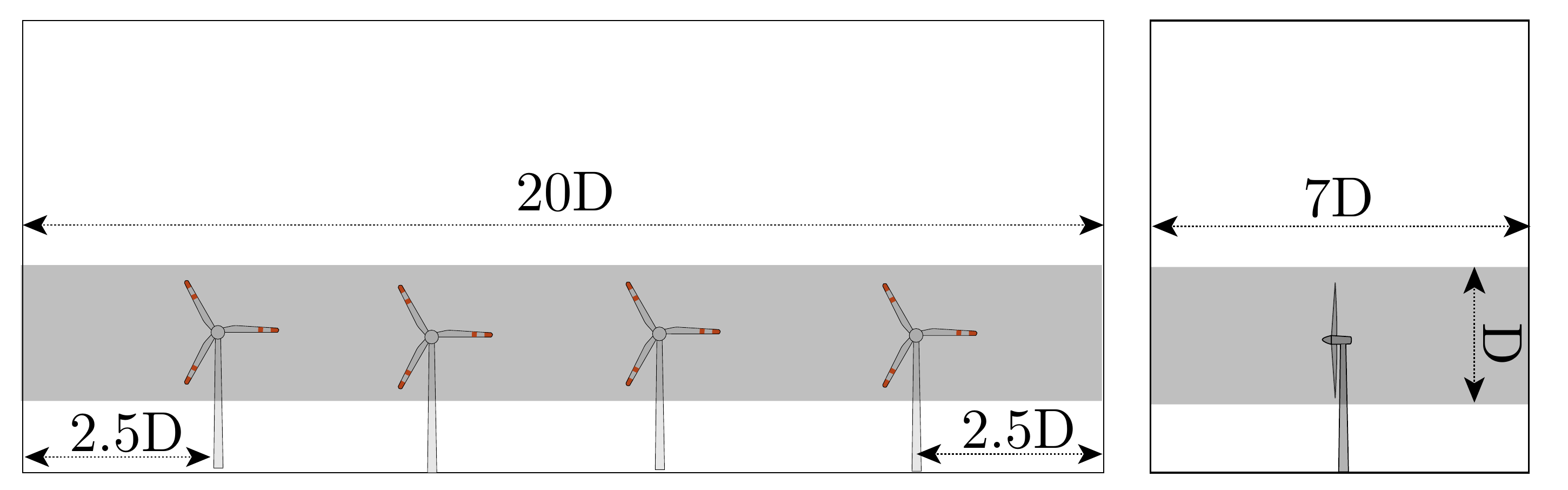}
 \end{subfigure}
 \caption{Shaded area represents the control volume used in the budget analysis. The control volume for each column has a dimension of $7D\times{20D}\times{D}$, and starts at a height of $z_h-D/2$.}
 \label{fig7}
\end{figure}

The steady-state, filtered energy equation is obtained by operating the momentum equation with $\widetilde{u}_i$ and performing time averaging \citep{all17, sag06}. The energy equation is,
%
%Differential equation
\begin{equation}
\centering
\begin{split}
 \overbrace{\overline{u}_j\partial_j\left({\frac{1}{2}\overline{u}_i\overline{u}_i}+\frac{1}{2}\overline{u'_iu'_i}\right)}^\text{Kinetic energy flux}+\overbrace{\partial_j\left(\frac{1}{2}{\overline{u'_ju'_iu'_i}}+\overline{u}_i\overline{u'_iu'_j}\right)}^\text{ Turbulent transport}+\overbrace{\partial_j\left( \overline{u_i\tau_{ij}}\right)}^\text{SGS transport}=\overbrace{-\partial_i{\left(\overline{pu_i}\right)}}^\text{Flow work}\\+\overbrace{g\beta(\overline{u_i\theta}-\overline{u}_i\theta_0)\delta_{i3}}^{\text{Buoyancy}}+\overbrace{f_c\left(\overline{u}_iU_g\right)\delta_{i2}-f_c\left(\overline{u}_iV_g\right)\delta_{i1}}^{\text{Geostrophic forcing}}+\overbrace{\overline{f_iu_i}}^{\text{Turbine power}}+\overbrace{\overline{\tau_{ij}S_{ij}}}^{\text{Dissipation}}, 
\end{split} \label{eqn4.1} 
\end{equation}
where, the overline represents time averaging, and $\overline{u'_iu'_j}=\left(\overline{{u_iu_j}}+\overline{\tau_\mathit{ij}}\right) - \overline{{u}_i}~\overline{{u}_j}$ represents the momentum flux to which the SGS components have been added. We are interested in the total power production per wind turbine row and energy balance around each turbine. To calculate the total energy, we numerically integrate the terms in equation \eqref{eqn4.1} in a control volume $\forall$ surrounding each turbine row. Figure \ref{fig7} schematically represents the dimensions and the extent of the aforementioned control volume. The control volume covers all the turbines in a row and has a streamwise extent of $s_x$D, i.e.\ $7D$, with $3.5$D in front and $3.5$D behind the turbines, in the streamwise direction \cite{all17}. The control volume has a dimension of D in the vertical direction and covers the volume between $z_h-D/2$ and $z_h+D/2$. In the spanwise direction, the control volume covers the whole row with an additional $2.5D$ on the sides, essentially $20D$. So the total control volume size for each row is, $7D\times{20}D\times{}D$. {\color{black} It is worth mentioning here that the ends of the computational domain in the spanwise direction are not included in the control volume and are therefore not shown in figure \ref{fig7}, i.e. the fringe layers are not included in the energy budget analysis.\\} Integrating equation \eqref{eqn4.1} and rearranging gives
%
%Integral energy equation 
\begin{equation}
\centering
\begin{split}
 \overbrace{\int^{}_{\forall}\overline{f_iu_i}d\forall}^{\text{$\mathbb{P}$, Turbine power}}&=\overbrace{\int^{}_{S}\overline{u}_j\left({\frac{1}{2}\overline{u}_i\overline{u}_i}+\frac{1}{2}\overline{u'_iu'_i}\right)dS_{i}}^\text{$\mathbb{E}_k$, Kinetic energy flux}+\overbrace{\int^{}_{S}\left(\frac{1}{2}{\overline{u'_ju'_iu'_i}}+\overline{u}_i\overline{u'_iu'_j}\right)dS_{i}}^\text{$\mathbb{T}_\text{t}$, Turbulent transport}+\overbrace{\int^{}_{S}\left( \overline{u_i\tau_{ij}}\right)dS_i}^{\text{$\mathbb{T_\text{sgs}}$, SGS transport}}\\&+\overbrace{\int^{}_{S}{\left(\overline{pu_i}\right)}dS_i}^{\text{$\mathbb{F}$, Flow work}}-\overbrace{\int^{}_{\forall}g\beta(\overline{u_i\theta}-\overline{u}_i\theta_0)\delta_{i3}d\forall}^{\text{$\mathbb{B}$, Buoyancy}}-\overbrace{\int^{}_{\forall}f_c\left(\overline{u}_iU_g\right)\delta_{i2}-f_c\left(\overline{u}_iV_g\right)\delta_{i1}d\forall}^{\text{$\mathbb{G}$, Geostrophic 
 forcing}}-\overbrace{\int^{}_{\forall}\overline{\tau_{ij}S_{ij}}d\forall.}^{\text{$\mathbb{D}$, Dissipation}}
\end{split}\label{eqn4.2}
\end{equation}
In equation \eqref{eqn4.2}, $\mathbb{E}_k$ represents the divergence of the kinetic energy flux, which includes both the resolved and the SGS kinetic energy, the turbulent transport term $\mathbb{T}_\text{t}$, which includes the entrainment of mean momentum due to turbulence and the entrainment of turbulent kinetic energy due to fluctuating velocities (third-order terms), $\mathbb{T}_\text{sgs}$ represents the transport of momentum due to SGS fluxes. The flow work $\mathbb{F}$ represents the energy transfer due to the static pressure drop of the flow across a turbine. The term $\mathbb{B}$ represents the turbulence destruction due to buoyancy, $\mathbb{G}$ represents the mean geostrophic forcing, and $\mathbb{P}$ represents the turbine power production.

\begin{figure}
 \centering
 \begin{subfigure}[ht!]{0.85\textwidth}
 \includegraphics[width=\linewidth]{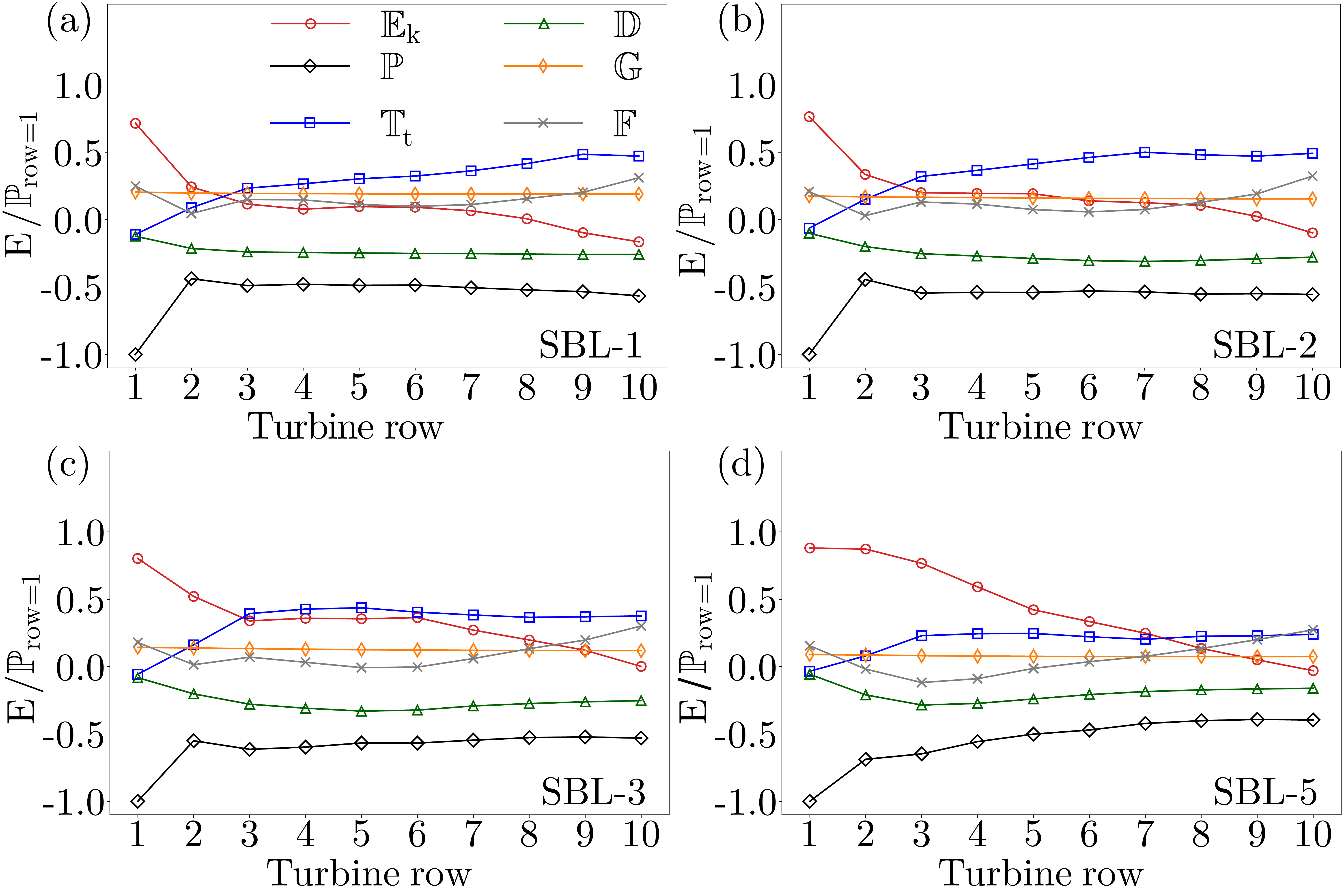}
 \end{subfigure}
 \caption{Energy budget for cases (a) SBL--1, (b) SBL--2, (c) SBL--3, and (d) SBL--5. All the terms are normalized by the power production of the first turbine row. The symbols in the legend are defined in equation \eqref{eqn4.2} and information about the cases can be found in table \ref{table1}.}
 \label{fig8}
\end{figure}

We are interested in the contribution of different budget components to power production. Therefore all the terms are normalized by the magnitude of the power produced by the first turbine row. The SGS transport $\mathbb{T}_\text{sgs}$ and the buoyancy fluxes $\mathbb{B}$ are small, less than $10\%$ of the first-column power and have been left out of the plots for brevity. The terms in equation \eqref{eqn4.2}, which include the gradients i.e.\ $\mathbb{E}_\text{k}$, $\mathbb{T}_\text{t}$, $\mathbb{T}_\text{sgs}$, and $\mathbb{F}$, represent the net flux out of the control volume, for example, $\mathbb{E}_\text{k} = \mathbb{E}_\text{out} - \mathbb{E}_\text{in}$. Positive values of these terms $\mathbb{E}_\text{in} > \mathbb{E}_\text{out}$ indicate that more energy is added to the control volume than removed. This indicates that in the control volume energy is extracted from the flow by the turbines or other means. Negative values of these terms indicate $\mathbb{E}_\text{out} > \mathbb{E}_\text{in}$, which means energy is being added to the flow.

For all the cases, the geostrophic forcing term $\mathbb{G}$ remains nearly constant for all the rows of the wind farm, representing a constant driving force. Besides $\mathbb{G}$, there are three primary energy sources, which determine the turbine power production, namely (i) the kinetic energy flux $\mathbb{E}_\text{k}$, (ii) the work done due to the static pressure drop $\mathbb{F}$, and (iii) the turbulent transport $\mathbb{T}_\text{t}$, which includes both entrainment of mean momentum into the wind farm by turbulent fluxes (shear production term) and the entrainment due to turbulent fluxes (third-order turbulence terms). Major energy sinks are the power extracted by the turbines $\mathbb{P}$, the dissipation $\mathbb{D}$, and the turbulence destruction due to buoyancy $\mathbb{B}$.

\begin{figure}
 \centering
 \begin{subfigure}[ht!]{0.75\textwidth}
 \includegraphics[width=\linewidth]{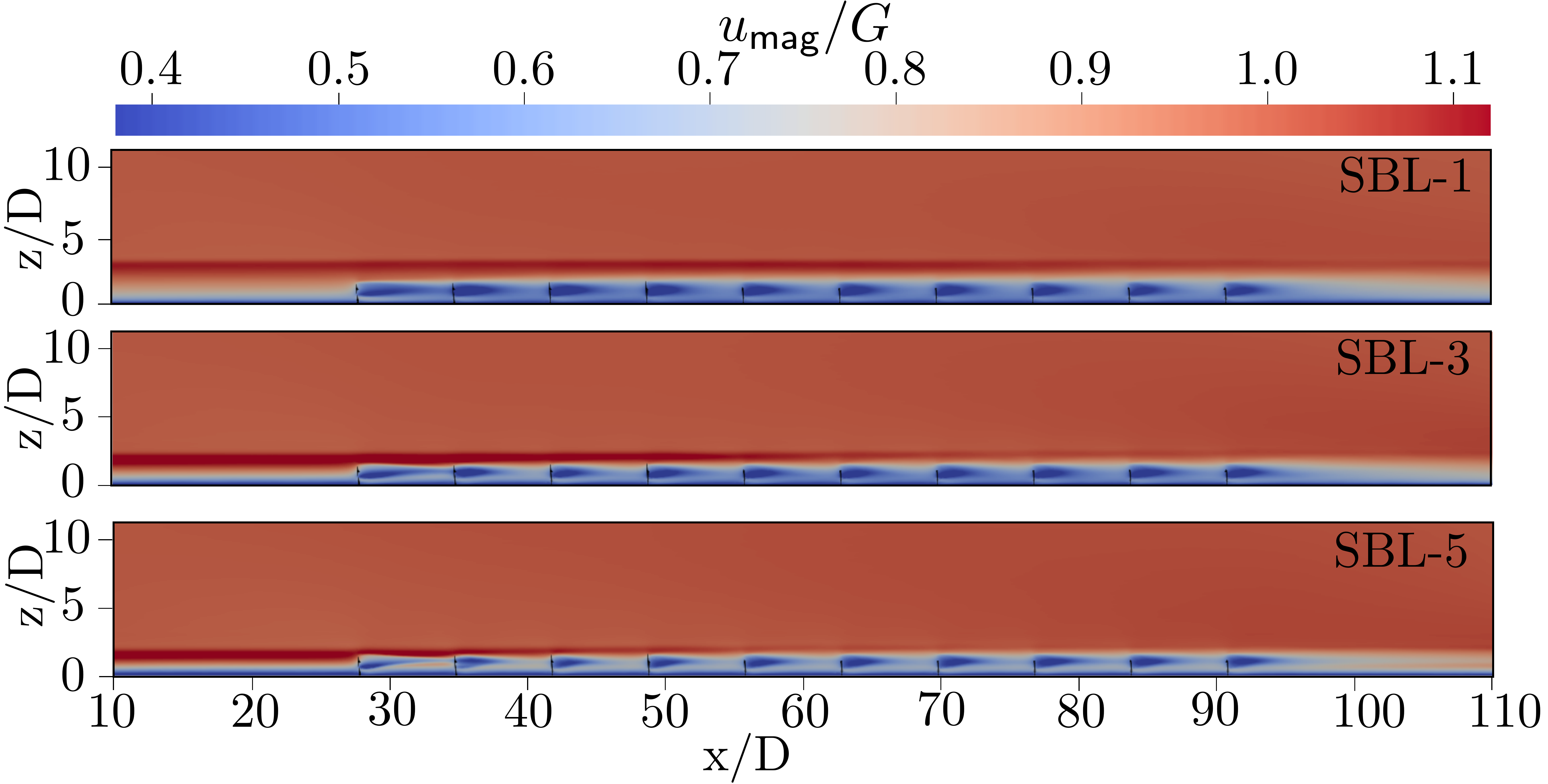}
 \end{subfigure}
 \caption{Streamwise velocity contour for different cases. Note that the strength of the LLJ is negligible towards the rear for the wind farm for SBL--5.}
 \label{fig9}
\end{figure}

Figure \ref{fig8}(a) shows different energy components for the SBL--1 case. The turbines continuously extract energy from the flow, and the kinetic energy flux decreases in the downwind direction. {\color{black} Furthermore, $\mathbb{E}_\text{k}$ is composed of two components, a mean energy component and a turbulent component. The mean component is directly related to the mechanical shear of the LLJ, while the fluctuating component is due to turbulence.} For the last three rows, $\mathbb{E}_\text{k} < 0$, which means more energy leaves the control volume than enters it. This happens because of the entrainment of the kinetic energy $\mathbb{T}_t$ from above the wind farm. The turbulent transport term $\mathbb{T}_\text{t}$ is composed of fluxes like $\overline{u}\cdot\overline{u'w'}$ and $\overline{v}\cdot\overline{v'w'}$, which represent the vertical (downward) flux of the mean momentum created by turbulence, i.e.\ entrainment of mean energy from above towards the turbines. The entrainment flux increases in the downwind direction due to the increased turbulence levels created by the wind turbine wakes. In a wind farm operating under neutral stratification and no LLJ, this entrainment flux is of the same order of magnitude as the turbine power production. This flux acts as the major source of power for the downwind wind turbines and reaches a constant value towards the end of the wind farm \citep{cal10, cal10b}. A similar variation of energy fluxes has been reported in the simulations of wind farms in conventionally neutral boundary layers \cite{all17}. For SBL--1, the jet height ($z_{jet}/z_h=2.670$) is well above the wind farm. The IBL grows above the wind farm and facilitates the interaction with the high-velocity jet. Consequently, the entrainment continuously increases downwind and reaches its maximum towards the end of the wind farm. Figure \ref{fig9} shows that although the jet strength reduces for SBL--1, the jet more or less persists above the entire wind farm. Figure \ref{fig8}(a) shows that the pressure-velocity correlation due to the static pressure drop, also known as the flow work $\mathbb{F}$, is positive and increases along the length of the wind farm. This indicates that the turbines operate in a favorable pressure gradient in the SBL--1 case. $\mathbb{F}$ has a significant contribution towards the power production near the end of the wind farm. The turbine power production, which is the major sink, is maximum at the entrance and reduces downwind due to the effect of the upwind turbine wakes. This variation is typical for a wind farm with an aligned layout and has been observed in field measurements and numerical studies \citep{han12, ste14b, wu13}. The dissipation $\mathbb{D}$ acts as an additional energy sink and remains roughly constant as a function of the downwind position in the wind farm.\\
%%%%%%%%%%%%%%%%%%%%%%%%%%%%%%%%%%%%%%%%%%%%%%%
%%%%%%%%%%%%%%%%%%%%%%%%%%%%%%%%%%%%%%%%%%%%%%%%%%
\begin{figure}
 \centering
 \begin{subfigure}[ht!]{0.85\textwidth}
 \includegraphics[width=\linewidth]{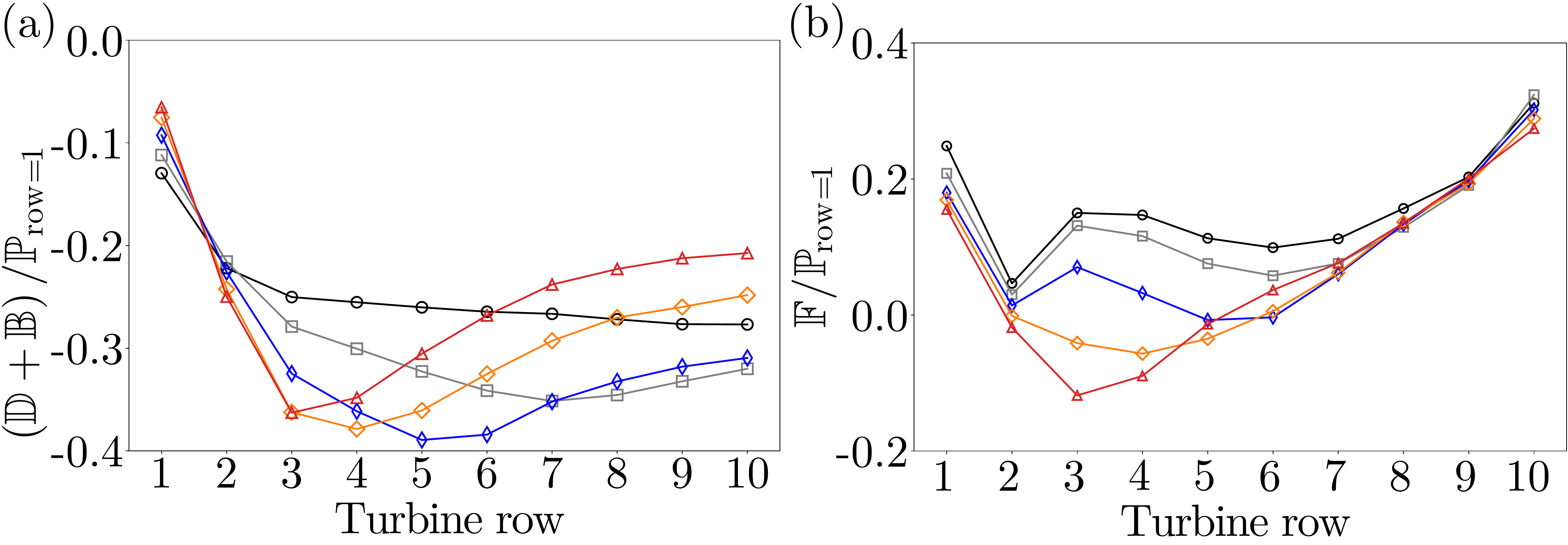}
 \end{subfigure}
 \caption{(a) Turbulence destruction due to buoyancy and dissipation, $\mathbb{B}+\mathbb{D}$ (b) Flow work $\mathbb{F}$ represents the work done due to pressure drop.}
 \label{fig10}
\end{figure}
%%%%%%%%%%%%%%%%%%%%%%%%%%%%%%%%%%%%%%%%%%%%%%%%%%
\indent Figure \ref{fig8}(b) presents the energy budget for the SBL--2 case. The figure shows that the entrainment $\mathbb{T}_\text{t}$ increases until the {\color{black}7$^\text{th}$ row} when it saturates. A similar trend is observed for the SBL--3 case in figure \ref{fig8}(c), but then the entrainment already saturates after the {\color{black}5$^\text{th}$ row}. For SBL--3 the jet height ($z_\text{jet}/z_h=1.836$) is slightly above the turbine tip height. The increase and decrease in entrainment correspond to the positions when the wind farm IBL starts interacting with the LLJ. Figure \ref{fig9} shows that the jet strength for SBL--3 is significantly reduced after the {\color{black}5$^\text{th}$ row}. For SBL--5, the jet is utilized by a couple of rows at the entrance, and the remaining rows have little or no jet left to entrain, therefore $\mathbb{T}_\text{t}$ remains nearly constant for this case after the initial increase.\\
\indent Figure \ref{fig10}(a) shows the variation of $\mathbb{D} + \mathbb{B}$ for different cases. Both $\mathbb{B}$ and $\mathbb{D}$ act as energy sinks in the budget, and the buoyancy flux $\mathbb{B}$ is small, i.e.\ less than 8\% of the first-row power for all the cases. Therefore $\mathbb{B}$ is combined with $\mathbb{D}$ to represent the net energy sink. $\mathbb{B}+\mathbb{D}$ is maximum when the turbines interact with the LLJ. This shows that the turbulence production due to mean shear is maximum when the LLJ is at lower heights. In SBL--5, for which the stability is the highest (see figure \ref{fig8}(d)), $\mathbb{T}_\text{t}$ is nearly equal to $\mathbb{D}$, which means there is no effect of entrainment fluxes on the turbine power production and we see a continuous drop in the kinetic energy flux as well as power production. Under stable stratification, increasing the stability damps out the vertical velocity fluctuations, which results in a reduction of in the downward transport of horizontal momentum towards the surface (see figure \ref{fig3}(d)). This results in a reduction of shear production terms $\overline{u'w'}\partial{\overline{u}}/\partial{z}$ and $\overline{v'w'}\partial{\overline{v}}/\partial{z}$ in $\mathbb{T}_\text{t}$, which causes a reduction of the turbulent kinetic energy. As mentioned before, the absolute value of $\mathbb{B}$ is not significant. However, the turbulent fluctuations damped out by the stratification, in turn, affect the momentum flux, which causes the weak turbulence in the SBL \citep{sha14}. With the jet utilized by the first {\color{black}few turbine rows in SBL--5}, the turbines downwind experience a reduction in shear production and mean shear. Consequently, we see a continuous decrease in power production for the turbines further downwind.\\
\indent {\color{black}Monin and Yaglom (1971) \cite{mon71}} describe the Obukhov length as the height below which buoyancy or the thermal effects do not play an important role. In a SBL, for $z<<|L|$, the effects of dynamic factors such as shear dominate. For $z>|L|$ the thermal effects dominate diminishing turbulence. The Obukhov length for cases SBL--5 and SBL--4 are 48.8 m and 66.0 m, respectively, which is less than the turbine hub height. In these cases, the turbines operate mostly in a buoyancy dominated region with high stability. Therefore, we see minimal shear production and turbulent transport $\mathbb{T}_\text{t}$ in these cases. Here, $\mathbb{T}_\text{t}$ is more or less balanced by $\mathbb{B}+\mathbb{D}$ (figure \ref{fig8}(d)), and the turbine power production depends completely on non-turbulent phenomena such as the divergence of mean kinetic energy flux and the static pressure drop. With the increased shear associated with LLJ, the turbines in a SBL produce more power than the turbines operating in the absence of a LLJ. For cases with high stability i.e.\ $z_h<|L|$, $\mathbb{E}_\text{k}$, $\mathbb{F}$, and $\mathbb{G}$ are the only energy sources available, as $\mathbb{T}_\text{t}$ is balanced by $\mathbb{B}+\mathbb{D}$. Therefore, the power production decreases with increasing stratification. However, even in the presence of a LLJ the front turbine rows may perform well due to the elevated shear in the LLJ.\\
\indent Figure \ref{fig10}(b) presents the variation of the flow work $\mathbb{F}$ for different cases. SBL--1, SBL--2, and SBL--3 show that the flow work is always positive, which shows that the turbines operate under a favorable pressure gradient. Since $\mathbb{F}>0$, it acts as an energy source for the turbine power production for the cases SBL--1, SBL--2, and SBL--3. For the case SBL--5, with the increase in streamwise distance, the resistance to the flow created by the surface inversion top increases as the IBL grows. This resistance to the flow reaches a maximum at the third turbine column (approximately $x/D\approx45$) when the IBL height is the same as the height of the inversion top, and we see the minimum of $\mathbb{F}$ at this point. Following this critical point, the flow starts going around the wind farm, and consequently the pressure drop across the wind farm increases.
%%%%%%%%%%%%%%%%%%%%%%%%%%%%%%%%%%%%%%%%%%%%%%%%%%%%%%%%%%%%%%%%%%%%%%%%%%%%%%%%%%%%%%
\subsection{Turbine power production}\label{sec4.2}
\begin{figure}
 \centering
 \begin{subfigure}[ht!]{0.85\textwidth}
 \includegraphics[width=\linewidth]{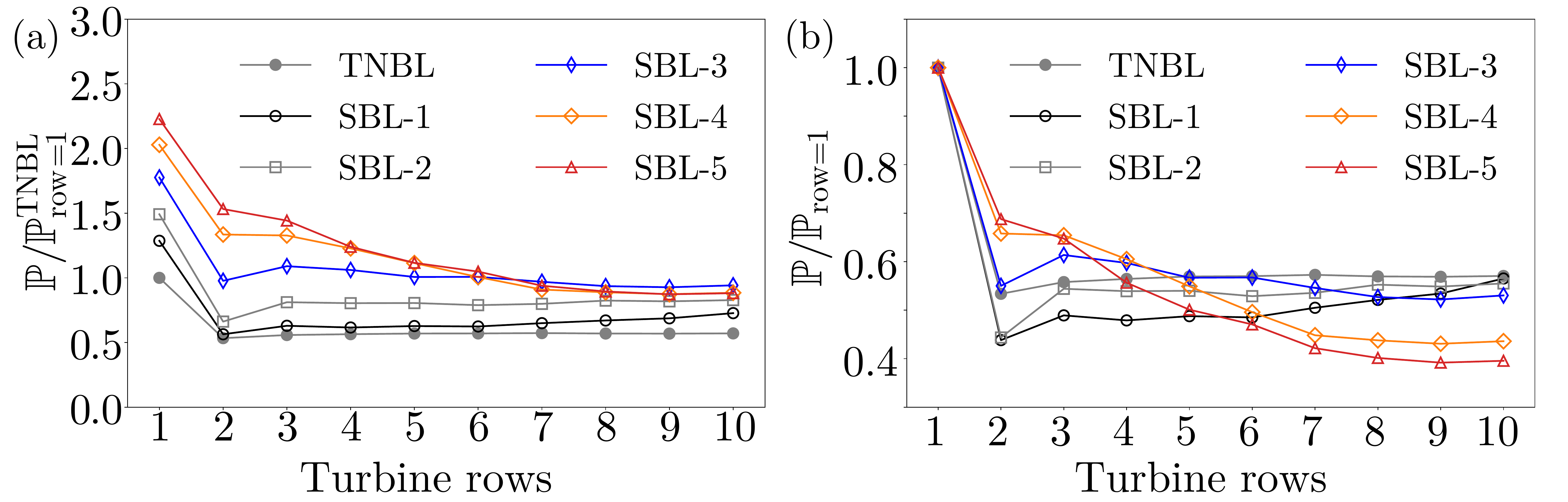} \end{subfigure}
 \caption{(a) Power production normalized with the first-row average power of SBL--1. (b) Power production normalized with the power production of the first-row.}
 \label{fig11}
\end{figure}

%%%%%%%%%%%%%%%%%%%%%%%%%%%%%%%%%%%%%%%%%%%%%%%%%%%%%%%%%%%%%%%%%%%%%%%%%%%%%%%%%%%%%%
{\color{black} Figure \ref{fig11}(a) presents the power production of different cases normalized by the first-row power production of the TNBL case. The figure shows that turbines in the presence of a jet produce more power than in a TNBL. As mentioned previously, we used a friction velocity of $0.316$ m/s obtained from the SBL--1 case for the TNBL case. The figure also shows that the power production of the first turbine row increases significantly when the surface cooling is increased. The reason is that the average hub height velocity is higher for the cases with stronger stratification, see figure \ref{fig3}(a). However, the figure shows that the turbine power production towards the end of the wind farm is lower for cases SBL--4 and SBL--5 than for SBL--3. The reason is that the turbulent energy entrainment further downwind in the wind farm is limited for these cases. It is also worth mentioning here that in the presence of an `infinitely' wide turbine array, the induction region in front of the wind farm is more pronounced. Therefore, a `finite' wind farm produces more power than an `infinitely' wide wind farm.\\ 
\indent To study the effect of wake recovery on the performance of downwind turbine rows for the different cases, figure \ref{fig11}(b) presents the row-averaged power normalized by the first row power production. After the second row, an increase in power production indicates is a result of relatively fast wake recovery due to high turbulence, and a continuous decrease in power indicates slower wake recovery. For SBL--1 the $\mathbb{P}/\mathbb{P}_\text{row=1}$ increases downwind of the first turbine. This increase in the relative power production with the downwind direction indicates that more energy is entrained from the jet, which is then extracted by the turbines. For SBL--4 and SBL--5 $\mathbb{P}/\mathbb{P}_\text{row=1}$ decreases asymptotically to a constant value indicating reduced relative wake recovery. Furthermore, for the SBL--3, SBL--4, and SBL--5 cases, the wake recovery up to the fifth row is better than for the TNBL case. This is due to the lower-height of the LLJ. At low LLJ heights, the turbines can directly interact with the LLJ by wake meandering, leading to higher relative power for the first few rows. Further downwind, the wakes in neutral condition show better recovery than the stable cases due to higher turbulence intensity. The TNBL has higher relative production further downwind because the turbulence intensity, which is the dominating factor for wake recovery, is higher in a neutral boundary layer than under stable stratification.} \\
\indent {\color{black} We find that the turbine power fluctuations decrease with increasing stability (not shown here). This is in agreement with the decrease of the atmospheric turbulence intensity with increasing thermal stratification. Downwind of the first turbine row, the fluctuations mainly depend on the wake generated turbulence. Tobin et al. \cite{tob19} report that the wake motions increase the turbine power fluctuations. We also observe an increase in the turbine power fluctuations of the downwind turbine rows due to the upwind turbine wakes (not shown here). This increase in the power fluctuations, even at higher stability, is due to the wake motions and increases wake recovery.}\\
\begin{figure}
 \centering
 \begin{subfigure}[ht!]{0.85\textwidth}
 \includegraphics[width=\linewidth]{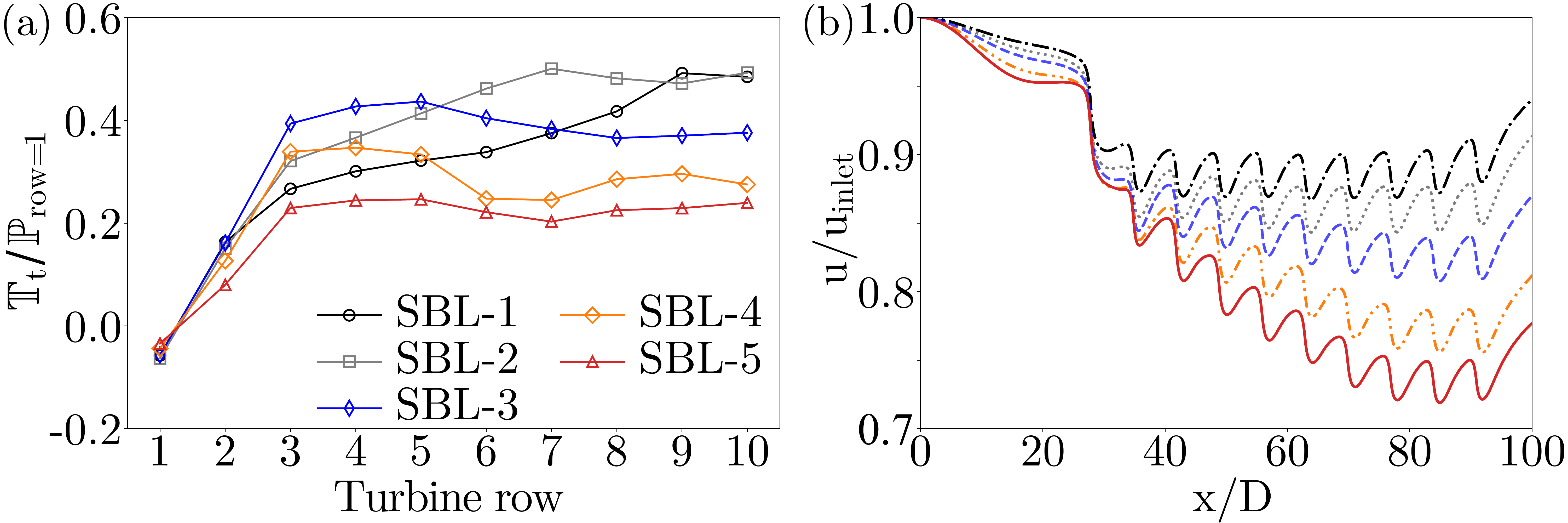}
 \end{subfigure}
 \caption{(a) The development of the turbulent transport term $\mathbb{T}_\text{t}$, see equation \eqref{eqn4.2}, as function of the downwind position. (b) Visualization of the streamwise velocity development at the hub height normalized by the inlet velocity.}
 \label{fig12}
\end{figure}
\indent Figure \ref{fig12}(a) and (b) show the turbulent entrainment and wake recovery for different stable cases. In the region behind the fifth row, the SBL--3 case shows maximum entrainment. In this case the jet height is $z_\text{jet}/z_h=1.836$, and due to the vertical meandering of the turbine wakes high-velocity wind from the jet is entrained. This interaction reaches a maximum around the 3$^\text{rd}$ turbine row, after which the jet is completely used up and the entrainment continuously decreases. Figure \ref{fig12}(b) shows that SBL--1 has the fastest wake recovery of all the cases. The inlet turbulence intensity at hub height for this case is the highest at $\mathrm{TI}_\mathsf{z_h}=5.82\%$. Cases SBL--2 and SBL--3 show significant wake recovery towards the end of the wind farm. For these cases the inlet Obukhov length is 189 and 100 meters, respectively, which is greater than the hub height. This means that the turbines are in a regime where there the shear generated turbulence effects dominate. As a result of the turbulence generated towards the end of the wind farm, these cases show significant wake recovery.
Figure \ref{fig12}(b) shows a significant reduction in the upwind wind velocity in front of the first turbine row, which indicates the effect of the adverse pressure gradient created by the wind farm blockage. This upwind reduction in wind speed increases with stratification and is highest for SBL--5 for which the adverse pressure gradient caused by the inversion is maximum. This flow blockage reduces the inlet wind velocity for the first row of turbines, and the turbines produce lesser power than what they would if they were free-standing. Similar upwind flow reduction has been observed in previous studies of wind farm flow blockage \cite{seg20, ble18, all18, wu17}.
%%%%%%%%%%%%%%%%%%%%%%%%%%%%%%%%%%%%%%%%%%%%%%%%%%%%%%%%%%%%%%%%%%%%%%%%%%%%%%%%%%%%%%
\section{Effect of wind veer}\label{sec5}
%%%%%%%%%%%%%%%%%%%%%%%%%%%%%%%%%%%%%%%%%%%%%%%%%%%%%%%%%%%%%%%%%%%%%%%%%%%%%%%%%%%%%%
\begin{figure}
 \centering
 \begin{subfigure}[ht!]{0.75\textwidth}
 \includegraphics[width=\linewidth]{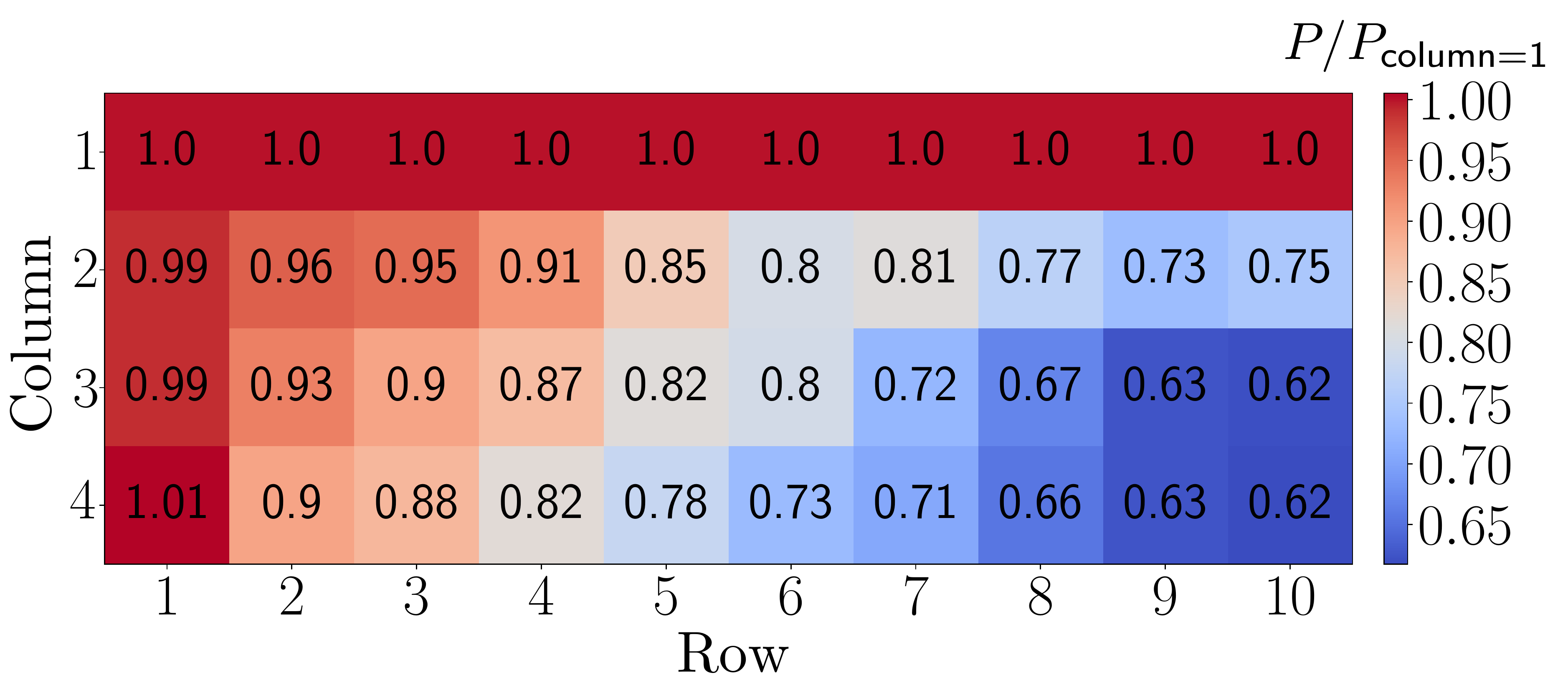}
 \end{subfigure}
 \caption{Power map for the case SBL--3. All the entries have been normalized by the power of the first column. Due to the wind veer, the first column produces more power compared to the other columns.}
 \label{fig13}
\end{figure}
\vspace{-4mm}
%%%%%%%%%%%%%%%%%%%%%%%%%%%%%%%%%%%%%%%%%%%%%%%%%%%
\begin{figure}
 \centering
 \begin{subfigure}[ht!]{\textwidth}
 \includegraphics[width=0.8\linewidth]{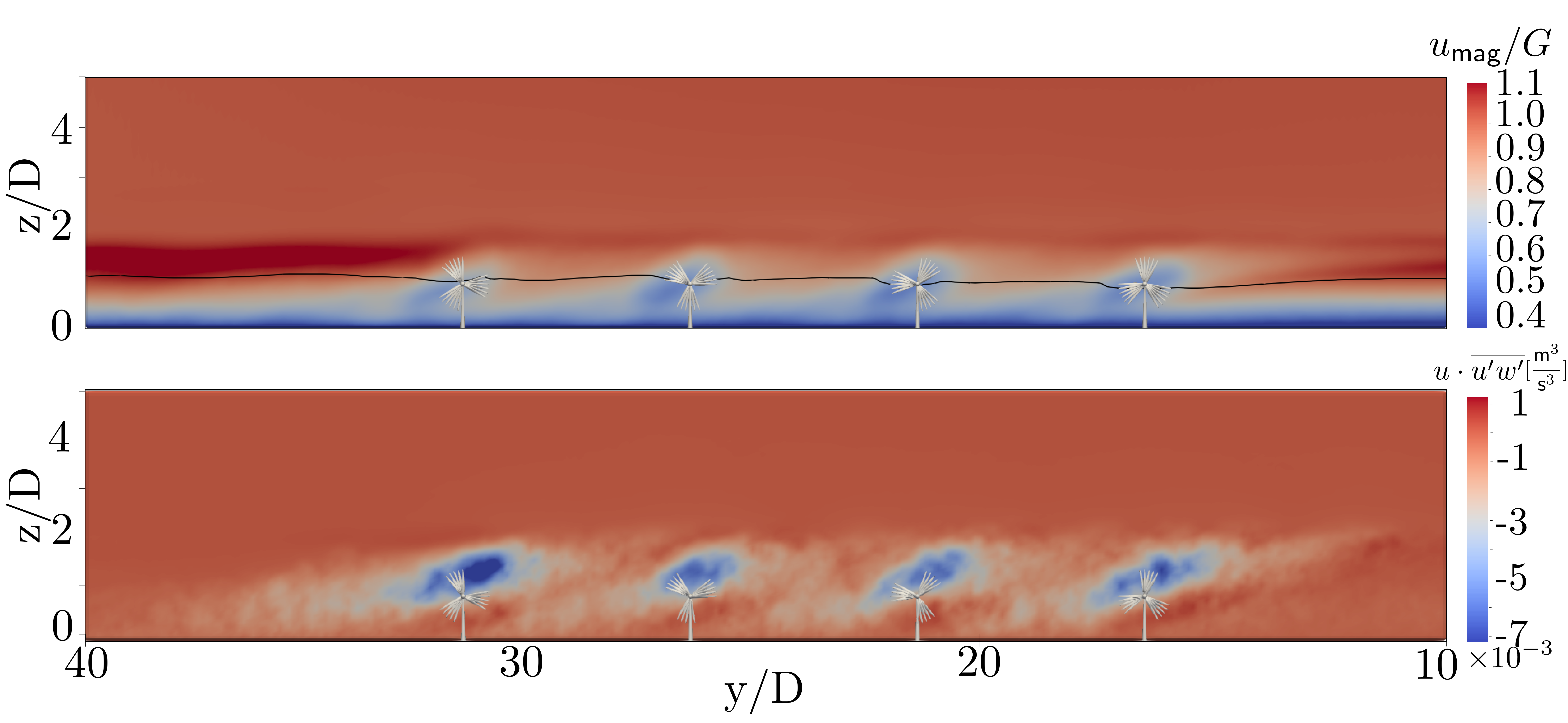}
 \end{subfigure}
 \caption{(a) Normalized horizontal velocity magnitude $u_{mag}/G$ and (b) the energy flux $\overline{u}\cdot\overline{u'w'}$ in the y-z plane, passing through the 6$^{th}$ turbine row. The black line represents the surface with zero spanwise velocity ($v=0$). Above the line, the flow goes to the right, and below the line, the flow is going to the left.}
 \label{fig14}
\end{figure}
%%%%%%%%%%%%%%%%%%%%%%%%%%%%%%%%%%%%%%%%%%%%%%%%%%%%%
In the presence of the Coriolis force, the wind follows an Ekman spiral, i.e. the wind velocity vector changes its direction with height. The changes in the wind angle are caused by the imbalance between the pressure gradient and frictional forces. Under stable stratification, the wind veer is very pronounced. In our simulations, we use a PI controller to fix the wind angle at the hub height to zero \cite{nag19}. This results in a flow that has a positive spanwise velocity below the turbine hub and a negative spanwise velocity above the turbine hub. The flow is turned such that the natural wind veer leads to these velocities in the frame of reference that we pick. \\
%%%%%%%%%%%%%%%%%%%%%%%%%%%%%%%%%%%%%%%%%%
\indent Figure \ref{fig13} presents the power map for the SBL--3 case with all the entries normalized by the power produced by the turbines in the 1$^\text{st}$ column of their respective rows. It is evident from the figure that the turbines in the 1$^\text{st}$ column produce more power compared to the rest of the turbine columns. Furthermore, there is a gradual reduction in power production towards the fourth column. This variation in power is because of the wind veer created by the Coriolis force. We find that this effect is substantial for SBL--3, SBL--4, and SBL--5. The effect is certainly present for SBL--1 and SBL--2 but not significant.\\
%%%%%%%%%%%%%%%%%%%%%%%%%%%%%%%%%%%%%%%%%%%
\indent Figure \ref{fig14}(a) shows the horizontal velocity magnitude for the SBL--3 case in the $y-z$ plane cut through the middle of the sixth turbine row. In this case, the jet height is $z_\text{jet}/z_h=1.836$, which is slightly above the turbines. We observe that the turbines completely utilize the jet above the wind farm due to entrainment and wake meandering, whereas the jet to the left of the first turbine column provides a continuous supply of fresh momentum due to the spanwise flow, which goes to the right. The turbines on the left in figure \ref{fig14} get a constant energy supply from the high-speed jet, which is utilized by the turbines, while the remaining fluid goes to the turbines on the right. As the first column has already utilized the jet, the power production of the next column is reduced. Furthermore, the local variation in the wind velocity created by the turbine wakes also causes the wind to deflect clockwise. The deflection of the turbine wakes clockwise in the Northern hemisphere is due to the imbalance created by the entrainment fluxes induced by the wind farm \cite{van17b, nag19}. We observe a similar clockwise deflection of the turbine wakes due to which the turbines in the inner columns operate in the wake of the outer columns.\\
\indent Figure \ref{fig14}(b) shows the streamwise downward energy flux $\overline{u}\cdot\overline{u'w'}$ for the SBL--3 case. The wake structure is skewed due to the lateral shear created by the spanwise flow. The turbine in the first column entrains most energy from the jet, and the subsequent columns entrain less energy from the jet due to the wind turbine wake. This skewed spatial structure of energy entrainment is an additional reason for the observed power variation. 
%%%%%%%%%%%%%%%%%%%%%%%%%%%%%%%%%%
\section{Conclusions}\label{sec6}
\indent We performed large-eddy simulations of wind farms in stable boundary layers. The objective of the study was two-fold: 1) to study the variation of wind farm power production with the LLJ height and 2) to study the effect of stable stratification on the flow development in wind farms. The study was carried out by systematically increasing the cooling rate at the surface , which results in lower LLJ height and a reduction of the atmospheric turbulence. At lower stratification, when the top of the surface inversion is significantly above the IBL height. In this case, the wind farm IBL is below the the top of the stable boundary layer and the flow accelerates over the wind farm. With increasing stratification, the boundary-layer height reduces, the fluid has less space to accelerate over the wind farm, and the flow goes around the wind farm. Therefore, performing simulations with periodic boundary conditions in the spanwise direction over-predicts the flow blockage as the flow cannot go around the wind farm.\\
%%%%%%%%%%%%%%%%%%%%%%%%%%%%%%%%%%%%%%%%%%%
\indent {\color{black} 
A wind farm interacts with a LLJ in two ways, firstly by wake meandering with low-height LLJs and secondly turbulent entrainment with LLJs high above. We find that power production of the first row increases when the LLJ height decreases. In addition, we find that the first-row power production is higher in the presence of a LLJ than for the reference case with neutral stratification without an LLJ, i.e. the TNBL case. Compared to weakly stable cases (SBL--1 \& SBL--2), TNBL case shows faster wake recovery due to high turbulence intensity. However, as long as energy can be entrained from the jet, the wake recovery for the stable boundary layers can be faster than for the TNBL case.} We observe increased entrainment when the jet is above the wind farm. The entrainment is strongest when the wakes can directly interact with the jet by the vertical meandering of the wakes. If the LLJs are at a height $z_\text{jet}\leq{z_h+D/2}$, the turbines at the entrance which can directly extract energy from the LLJ perform significantly better than the inner turbines. Under similar stability conditions, a wind farm performs better if the LLJ is present above the wind farm than when an LLJ is absent. The simulations show that the turbine rows at the entrance utilize the LLJ, and the entrainment decreases after the jet strength is reduced. Therefore, at sites where LLJs are prominent, wind farms with higher aspect ratios (spanwise width-to-streamwise length ratio of the wind farm) are beneficial over long wind farms with low aspect ratios.\\
\indent Stable atmospheric boundary layers generally have low turbulence intensities, and the surface Obukhov length can serve as an important length scale to predict the impact of the stability. We find that for $z_h>>|L|$ the shear effects dominate, and the entrainment is more than the dissipation and buoyancy destruction. When $z_h<|L|$ the thermal effects dominate, and there is very little entrainment as buoyancy damps out the vertical velocity fluctuations reducing both vertical kinetic energy and downward turbulent fluxes.\\
\indent In the presence of an LLJ, an appreciable spanwise flow is created by the wind veer. Consequently, the turbines which can directly interact with the LLJ (e.g.\ turbines in the left column in figure \ref{fig14}) produce more power than the rest of the turbines. The rest of the turbines can only interact with the LLJ via turbulent entrainment. This effect is prominent when the jet height $z_\text{jet} \approx z_h + D$. Finally, the present study only focuses on the cases where the jet is above the turbine top height, i.e.\ $z_\text{jet} \geq z_h+D/2$. Consequently, the turbines only experience positive shear in the LLJ. Further studies are required to analyze the effect of negative shear of the LLJ (when $z_\text{jet}<z_h +D/2$) on the wind farm power production.\\
\\
{\it Acknowledgements:} This work is part of the Shell-NWO/FOM-initiative Computational sciences for energy research of Shell and Chemical Sciences, Earth and Live Sciences, Physical Sciences, FOM, and STW. This work was carried out on the national e-infrastructure of SURFsara, a subsidiary of SURF corporation, the collaborative ICT organization for Dutch education and research.

\section*{References}
\bibliography{literature_windfarms}

\end{document}